\documentclass[12pt]{JHEP3}
\pdfoutput=1

\usepackage{amsmath,epsfig}
\usepackage{amssymb,amsfonts}
\usepackage{latexsym}
\usepackage[latin1]{inputenc}
\usepackage{tocvsec2}
\usepackage{subeqnarray}
\usepackage{xcolor}

\usepackage{graphicx}
\usepackage{longtable}

\relax
\renewcommand{\theequation}{\arabic{section}.\arabic{equation}}
\def\be{\begin{equation}}
\def\ee{\end{equation}}
\def\bea{\begin{eqnarray}}
\def\eea{\end{eqnarray}}

\newcommand\fverb{\setbox\pippobox=\hbox\bgroup\verb}
\newcommand\fverbdo{\egroup\medskip\noindent%
                        \fbox{\unhbox\pippobox}\ }
\newcommand\fverbit{\egroup\item[\fbox{\unhbox\pippobox}]}

\newcommand{\bear}{\begin{eqnarray}}

\newcommand{\eear}{\end{eqnarray}}

\newcommand{\bsea}{\begin{subeqnarray}}
\newcommand{\esea}{\end{subeqnarray}}
\newbox\pippobox

\def\6{\partial}

\def\a{\alpha}
\def\g{\gamma}
\def\nn{\nonumber}

\def\m{\mu}
\def\n{\nu}

\def\s{\sigma}

\def\sp{\;\;\;,\;\;\;}

\def\sq
\def\a{\alpha}

\def\hri#1#2{\href{http://arxiv.org/abs/#1}{[ArXiv:#1]#2}}
\def\hre#1#2{\href{http://arxiv.org/abs/#1/#2}{[ArXiv:#1/#2]}}
\def\hspi#1#2{\href{http://www.slac.stanford.edu/spires/find/hep/www?irn=#1}{#2}}

\def\d{\delta}

\newcommand{\ud}{\mathrm{d}}

\def\nn{\nonumber}

\def\s{\sigma}

\newcommand{\comments}[1]{}
\allowdisplaybreaks[3]

\title{Holographic Metals and Insulators with Helical Symmetry}
\author{\large  Aristomenis Donos$^{a}$, Blaise Gout\'eraux$^{b}$ and Elias Kiritsis$^{c,d,e}$\\
~\\
$^a${DAMTP, University of Cambridge, Cambridge, CB3 0WA, U.K.}\\
~\\
$^b$\href{http://www.nordita.org}{Nordita}, KTH Royal Institute of Technology and Stockholm University\\
Roslagstullsbacken 23, SE-106 91 Stockholm, Sweden.\\
~\\
$^c$\href{http://www.apc.univ-paris7.fr}
{APC, Universit\'e Paris 7}, CNRS/IN2P3, CEA/IRFU, Obs. de Paris, Sorbonne Paris Cit\'e, B\^atiment Condorcet, F-75205, Paris Cedex 13, France (UMR du CNRS 7164).\\
~\\
$^d$\href{http://wwwth.cern.ch/}{Theory Group, Physics Department, CERN}, CH-1211, Geneva 23, Switzerland.\\
~\\
$^e$ \href{http://hep.physics.uoc.gr}{Crete Center for Theoretical Physics},
Department of Physics, University of Crete, 71003 Heraklion, Greece.
\\\\
E-mail: \email{a.donos@damtp.cam.ac.uk}, \email{blaise@kth.se}, \href{http://hep.physics.uoc.gr/~kiritsis/}{http://hep.physics.uoc.gr/~kiritsis/}
}

\preprint{CCTP-2014-6\\CCQCN-2014-26\\CERN-PH-TH/2014-035\\NORDITA-2014-76}

\abstract{Homogeneous, zero temperature scaling solutions with Bianchi VII spatial geometry are constructed in Einstein-Maxwell-Dilaton theory. They correspond to quantum critical saddle points with helical symmetry at finite density. Assuming $AdS_{5}$ UV asymptotics, the small frequency/(temperature) dependence of the AC/(DC) electric conductivity along the director of the helix are computed. A large class of insulating and conducting anisotropic phases is found, as well as isotropic, metallic phases. Conduction can be dominated by dissipation due to weak breaking of translation symmetry or by a quantum critical current.  }


\begin{document}

\section{Introduction, Summary of results and Outlook}

\subsection{Motivations}

Conductivity is one of the central observables of condensed matter systems. In standard examples it is used to classify materials in classes labeled metals, superconductors  and insulators, and depends in many cases non-trivially on the underlying interactions of the electronic component.
It is relatively easy to measure both in DC and AC contexts. The DC conductivity of a metal increases as the temperature is lowered, while it decreases for an insulator.

For Fermi liquids, few mysteries remain concerning their conductivity mechanisms.
For strongly-coupled materials like the high-$T_c$ superconductors and heavy fermions, conductivity has a non-trivial behavior, with a landmark linear form, and it is fair to say that its origin is not known from first principles.
In most of the phase diagrams there is convincing evidence that the underlying normal states are not Fermi liquids.   Moreover, in exotic cases, like the c-axis conductivity of cuprates, the frequency dependence remains largely a mystery.

Holographic approaches, originating in string theory and the AdS/CFT correspondence, have been used recently to analyze models that may be in the same universality class as strongly correlated electrons.
Such approaches work in a context of large-N adjoint theories at strong coupling. The adjoint nature of such theories makes the large-N limit intractable. In the strong coupling limit however, the theory can be solved using an appropriate gravitational dual theory in  higher space-time dimensions.

Such techniques and their applications to phenomena at finite density, relevant for condensed matter systems, have been analysed in the last few years and several new concepts have emerged, see \cite{H} for a recent review.

Most holographic systems analyzed at finite density are translationally invariant.\footnote{Exceptions also exist, using D-brane defects and magnetic vortices, \cite{def1}-\cite{def4}.} The standard symmetry argument then indicates that the real part of the AC conductivity will have a $\delta(\omega)$ contribution as in a translationally invariant system a constant electric field generates an infinite current. The $\delta$ function has been argued in \cite{Hartnoll:2009sz,Herzog:2009xv} to be related, via causality, to a ${1\over \omega}$ pole in the imaginary part of the conductivity.
This $\delta$ function is distinct from the one appearing in superfluid/superconducting phases.

In most holographic cases the translationally invariant systems are metals.
There have been also systems that have a gapped spectrum in the current-current correlator, and they are therefore candidates for insulators. 
At zero density such systems were described in \cite{witten,ihqcd}.\footnote{See \cite{Mefford:2014gia} for a recent example of a gapless, dissipationless insulator at zero density.}

More interestingly, at finite density a large class of systems were found,\footnote{One particular case was, descending from an M-theory compactification was found independently in \cite{mcg}.} \cite{cgkkm}, that has a gapped spectrum for the current correlator. The spectrum of excitations is discrete, but there is again a zero mode because of translational invariance. This phase seems to be unique to holography and is a hybrid between an insulator and a perfect metal. Under strict DC fields it conducts perfectly, but under AC ones it is generically insulating.

Interestingly, the thermodynamics for such systems is a bit like Yang-Mills.
At finite temperature, up to a transition temperature $T_c$, the dynamics is temperature-independent to leading order in the large-N limit. However there is a first-order transition at $T_c$ to a new phase that is conducting.

In all of the above, the saddle points are translationally invariant and the DC conductivity is strictly infinite because of the $\delta$-function.
This is however a problem when one is interested in comparing to realistic finite density systems. In such cases, there is always a breaking of translational invariance, due to the ionic lattice as well as other impurities that may exist in the system and in most cases, their interactions with the electron gas determine the DC conductivity, by controlling the rate of momentum dissipation of the electron gas.

 The interaction with momentum dissipation agents has been discussed in rather general terms in  \cite{impurities, Hartnoll:2012rj}. When the interaction with dissipators is IR irrelevant with respect to RG fixed points, a perturbative IR calculation can determine the scaling of the IR DC conductivity. When the dissipation is IR relevant, it can change the nature of the saddle point, turning the system into an insulator as was first argued in \cite{Donos:2012js} and demonstrated for a class of models.

There have been several lines of research addressing the breaking of translational invariance in holographic saddle points at finite density and its impact on conductivity.
A first line of research introduced a holographic lattice imprinted by boundary conditions on the bulk charge contribution, \cite{lattice1}-\cite{lattice4}.
The system is then governed by PDEs that can be solved so far only numerically. In the regimes accessible to the numerical calculations, the lattice perturbation is irrelevant in the IR and it controls to leading order in the IR the DC conductivity as predicted in \cite{Hartnoll:2012rj} on general principles. In \cite{lattice1} this has been done for RN black-hole type geometries (or more generally geometries that asymptote to those in the IR), while in \cite{lattice2}, \cite{lattice3} the analysis has been extended to semilocal hyperscaling violating geometries in the IR.

Another line of approach proposed first in \cite{Vegh} was to assume an effective action  treatment for momentum dissipation associated to the breaking of translational invariance. It is known, \cite{K,ACK} that when translational invariance broken, the graviton obtains a mass corresponding via the holographic correspondence  to the anomalous dimension of the appropriate stress tensor components. By introducing therefore a mass term in the effective gravitational description, one introduces a source for translational invariance breaking and associated momentum dissipation leading to a finite DC conductivity. It is worth noting that even though the concept of the dual field theory energy momentum tensor is not entirely clear in the context of massive gravity one can still study electric conductivities.

The massive graviton approach was extended in \cite{Davison,BT, Amoretti:2014zha} and provided for a formula for the DC conductivity in the context of massive gravity.
As applied so far, it has its own limitations. It is well-known (see \cite{K,ACK}) that the proper framework of momentum non-conservation is to consider separately the two stress-tensors that exchange momentum. In a fully holographic large N context this is modeled by two large-N QFTs that interact with each other, exchanging momentum by some interaction.\footnote{The interactions described in \cite{K,ACK}, namely $\int dtd^dx ~O_1(x,t)O_2(x,t)$ exchange both momentum and energy ($O_1(x,t)$, $O_2(x,t)$ are operators in respectively QFT$_1$ and QFT$_2$). One can obtain uniform momentum exchange only by replacing this interaction by
$\int dt_1 dt_2~d^dx ~O_1(x,t_1)O_2(x,t_2)$.} The geometric bulk picture is of two asymptotically AdS spaces interacting via correlated boundary conditions at their common boundary.  In this context, the graviton mass is of order ${\cal O}(1)$ while the kinetic terms are of order ${\cal O}(N^2)$.

One may also entertain the situation where one of the two theories has an $N\sim {\cal O}(1)$. In this case, the graviton mass is of order  ${\cal O}(N^2)$ and the geometrical picture changes. The small $N$ QFT lives at the boundary of the holographic one and the coupling is localized in the bulk. To our knowledge, only the previous setup was analyzed and the relevant RG flows were studied in great detail in \cite{KN}.

The effective approach used in \cite{Vegh}-\cite{Amoretti:2014zha} should be obtained as a limit where the momentum dissipating sector has a much larger number of degrees of freedom so that the backreaction of  the dissipated momentum can be neglected and its metric frozen. It will be interesting to work this out explicitly using the holographic picture and derive the effective setup used so far.

A related issue concerns  the graviton potential used in the studies \cite{Vegh}-\cite{Amoretti:2014zha}. This has been chosen to belong to a very restricted class of potentials, \cite{dGT}, that for a single massive graviton and no other degrees of freedom guarantee the nonlinear absence of the Boulware-Deser ghost, \cite{BD}. It is known however by now that in the presence of other fields, massive gravitons have many other ways to avoid the nonlinear BD ghost, if such interactions are well-tuned, \cite{Sc,KN}. This is the case in string theory and holographic setups, allowing for more general graviton potentials.

The formula obtained for the DC conductivity in these studies is a sum of two contributions:
\be
\sigma_{DC}=\sigma_{DC}^{pc}+\sigma_{DC}^{drag}
\label{1}\ee
The first, $\sigma_{DC}^{pc}$ has been interpreted, \cite{KO},  as a quantum critical pair creation contribution as it persists at zero charge density. For the RN black hole it is a constant proportional to the inverse of the bulk gauge coupling constant that counts the relative density of charge-carrying degrees of freedom to the neutral ones in the strongly-coupled plasma. More generally, at finite density, it can be interpreted as a contribution from the quantum critical sector, \cite{impurities}. More recently, it was realised in \cite{Donos:2014cya} that the first term is the electric conductivity in the absence of a heat current. The interpretation as a pair creation term is then natural since charged pairs are created with zero total momentum and therefore not contributing to a net matter flow.

The second contribution is due to the effects of dissipating momentum. When translation-breaking operators are irrelevant, the system is expected to be metallic and this term should give the leading contribution to the DC conductivity. Then, a description of momentum relaxation in terms of the memory matrix formalism is apposite, and shows that the conductivity takes a Drude-like form, though no quasi-particle description is assumed \cite{Hartnoll:2012rj}.

This general form of the DC conductivity was seen already pure metric backgrounds in \cite{KO} and was generalized to dilatonic backgrounds in \cite{cgkkm}. In both cases, as the gauge field action is the DBI action, (\ref{1}) is replaced by
\be
\sigma_{DC}=\sqrt{(\sigma_{DC}^{pc})^2+(\sigma_{DC}^{drag})^2}
\label{2}\ee
giving results compatible with (\ref{1}) in the regimes where pair creation or drag diffusion dominates the conductivity.
In general, we expect a nonlinear formula that reflects the bulk action of the gauge field. In the probe DBI cases the momentum dissipation is due to the fact that charge degrees of freedom are subleading compared to uncharged ones. In a sense, there is a momentum conserving $\delta$-function but its coefficient is hierarchically suppressed.

In \cite{cgkkm} it was observed based on (\ref{2}) that for running scalars other than the dilaton and in 2+1 boundary dimensions, the drag DC resistivity, when it dominates,  is proportional to the electronic entropy. This is a general property of strange metals where both the measured electronic entropy and resistivity   are linear in temperature. This  was extended in \cite{DSZ} to more general cases using the massive graviton theory, and most importantly provided a kinetic explanation for the correlation suggesting a more general validity.

The lattice and massive gravity approaches where put in contact, \cite{BTV},  where an infinitesimal lattice perturbation of a charged holographic system was considered and by analyzing the bulk equations mapped  to the massive gravity framework.

A technology to break translational invariance while maintaining the simplicity of ODEs and remaining in the $\mathcal{O}(N^{2})$ supergravity limit was introduced in \cite{Donos:2012js}. At a technical level, the black hole construction was very similar to that of \cite{helical} with the difference that translations where broken explicitly on the $AdS_{5}$ boundary in order to give rise to a lattice. The deformation was introduced through a vector boundary operator which was relevant and three dimensional Euclidean symmetry is restored in the UV. The simplification of these constructions carries on even when calculating transport properties.

The structures left invariant under Bianchi VII$_0$ symmetry are helices of a fixed director. Whilst homogeneous, the horizons of these black holes break translational invariance and momentum along the director disipates. In the framework of spontaneous symmetry breaking in holography, helical symmetry was first discussed in \cite{ooguri} (see also \cite{Donos:2011ff}). In the context of holography, a gravitational Ansatz made out of invariant 1-forms was given in \cite{kachru-bia} where a classification of solutions of various Bianchi types was carried out and which moreover, are scale invariant.

More recently, along similar lines, two more proposals appeared in \cite{DG2} and \cite{AW}. In \cite{DG2} the phase of UV relevant, charge neutral, complex scalars was used in order to break translational symmetry on the boundary. From the bulk point of view this construction has the interpretation of two neutral scalars, which can be rotated among each other, and on the boundary the theory is deformed by both of their dual operators with a sinusoidal pattern which is relatively shifted by a phase of $\pi/2k$. In this setup both a lattice amplitude and a period are part of the UV boundary data.

The construction of \cite{AW} used bulk fields which are perturbatively massless on $AdS$. In order to relax momentum, they were given a linear dependence on the spatial coordinates of the boundary. This deformation has the interesting property that apart from being UV marginal it was also IR relevant yielding a $AdS_{2}\times\mathbb{R}^{2}$ extremal horizon even in the absence of background charge. In this case the only deformation parameter is a slope introduced in the UV, setting one more scale apart from the chemical potential.

As with any gravitational semiclassical description, axions and the associated exact translational symmetry are approximations in string theory. It is well-known from many arguments and theorems with partial validity that there are no continuous global symmetries in string theory that are not space-time symmetries, (see for example, \cite{BDi,book,BS}).  This implies in particular that the axion translational symmetries are broken to discrete symmetries (known as duality symmetries) by various non-perturbative effects in string theory.

The prototype such mechanism is the breaking of SL(2,R) continuous symmetry of classical IIB supergravity in ten dimensions to $SL(2,Z)$ by D-instantons, \cite{GG}. Such effects will generate a nontrivial potential for the axion in gauged supergravity (they do not in standard Poincar\'e supergravity), which will be periodic to respect the unbroken discrete duality symmetries. The same phenomenon is generated in four-dimensional gauge theories by instantons.
However, if the scalar associated with such breaking is the dilaton as in N=4 sYM or YM, the potential is exponentially suppressed at large $N$ as $e^{-N}$, and therefore can be neglected for the purposes discussed here.

For other scalars/axions where the instantons are world-sheet instantons  the non-perturbative (in $\alpha'$) axion potential cannot be neglected.
Despite this, axions can give a reliable information on the conductivities. In the IR scaling regime, they seem to match (very qualitatively) the physics   associated with homogeneous disorder on transport discussed recently in \cite{SS}. To what extend this similarity goes beyond the scaling of conductivity remains to be explored.

When the massless scalars of \cite{AW} are coupled to additional dilatons or when the moduli of the complex scalars of \cite{DG2} are allowed to run unboundedly, the possible deep IR solutions look identical. This has been exploited in \cite{Donos:2014uba,g2014} in the context of EMD theories to provide a large class of IR solutions which exhibit varying characteristics when it comes to conductivity (which, as the action is linear in the gauge field kinetic term $F^2$, have a form similar to (\ref{1}), with the two components having a similar interpretation).\footnote{See \cite{Taylor:2014} for a similar analysis with non-canonical kinetic terms.} The ODE structure of the system enabled the possibility of having explicit formul\ae\ for the conductivity, including the Hall angle \cite{Blake:2014yla} and the full thermoelectric coefficient matrix \cite{Donos:2014cya}. The classes of ground states reached is comparable to the ones that will be described in this paper and indicates that these different classes are expanding the landscape of  theoretical ``holographic materials".\footnote{The universality classes described in this paper were announced before publication in \cite{DGK}.}

The purpose of the present work is to investigate the helical ground states in a system that is general enough to allow for a diverse landscape of IR behaviors that would extend those of \cite{Donos:2012js}. The main motivation for this are two-fold:
\begin{itemize}

\item To produce novel holographic models of insulators and metals with or without sharp Drude peaks,  that go well beyond what is known.

\item To potentially apply some of these findings to known strongly anisotropic systems with metallic behavior without sharp Drude peaks, like the cuprates along the $c$-axis, \cite{cooper}-\cite{zaanen} or other materials that might exhibit a helical symmetry.
\end{itemize}

\subsection{Setup}

In this paper we will consider EMD$_2$ theories with two U(1) gauge fields.\footnote{This is for convenience and simplicity. A similar analysis can be  done also with a single U(1) gauge field but it is more involved. It will be considered in a future work.} They correspond to two independently conserved U(1) charges in the dual QFT.
For applications to more realistic systems such charges could both be electric, but coming from sectors that at low temperature/energy do not exchange charge. Therefore one can consider that their individual charges are concerned.
We will also assume  5 bulk space-time dimensions, or 4 boundary dimensions, in order to obtain non-trivial states with helical symmetry.

A general $U(1)^2$ symmetric, two-derivative action, reads after redefinitions,
\be\label{Action3}
 S=M^{3}\int d^{5}x\sqrt{-g}\left[R-{1\over 2}(\partial\phi)^2+V(\phi)-\frac{Z_1(\phi)}4F_1^2-\frac{Z_2(\phi)}4F_2^2\right].
\ee
The action depends on three positive functions of the scalar field, $V,Z_{1,2}$. The IR dynamics, when the scalar is running to $\pm \infty$ (dilatonic solutions), as advocated in \cite{cgkkm}, is controlled by the asymptotics of the functions $V,Z_{1,2}$.

As usual, following the philosophy of \cite{cgkkm} we will assume that  the scalar functions asymptote  in the IR like\footnote{The notion of the IR limit is defined translated in field space. These are the asymptotics of the functions as $\phi\to\infty$ or $\phi\to -\infty$. They are relevant, if the scalar field flows to these values in the IR.}
\be\label{IRcouplings}
	V(\phi)\underset{IR}{\sim}V_0 e^{-\d\phi}\,,\quad Z_1(\phi)\underset{IR}{\sim}Z_{10}e^{\gamma_1\phi}\,,\quad Z_2(\phi)\underset{IR}{\sim}Z_{20}e^{\gamma_2\phi}\,.
\ee
Therefore, in the IR, the bulk action, apart from the dimensionful constants $ V_0,Z_{10,20}$ that affect simply the physics, depends on three important dimensionless parameters, $\delta, \gamma_{1,2}$. We set $Z_{10}=Z_{20}=1$ in the remainder of the paper.

We will consider solutions that are anisotropic in space, but have however a helical symmetry, known as Bianchi VII$_0$ symmetry. In a sense this symmetry group is a subgroup of the Euclidean symmetries of black holes with flat horizons.
They have been introduced and analysed in the context of EMD theories in \cite{kachru-bia,kasa}, as well as in \cite{helical, Donos:2012js} in theories without neutral scalars.
We postulate a helical Bianchi VII$_0$ Ansatz for our metric and gauge fields
\be\label{3}
	\ud s^2=-D(r)\ud t^2+B(r)\ud r^2+C_1(r)\omega_1^2+C_2(r)\omega_2^2+C_3(r)\omega_3^2
\ee
where we have introduced the Bianchi VII$_0$ left-invariant 1-forms
\be\label{4}
	\omega_1=\ud x_1\,,\quad \omega_2=\cos(kx_1)\ud x_2+\sin(kx_1)\ud x_3\,,\quad \omega_3=\sin(kx_1)\ud x_2-\cos(kx_1)\ud x_3\,.
\ee
The Ansatz for the scalar and the gauge fields is:
\be\label{5}
	\phi= \phi(r)\,,\quad A_1= A_1(r)\ud t\,,\quad A_2= A_2(r)\omega_2
\ee
so that $A_1$ only carries electric charge and $A_2$ is of magnetic type.
The Bianchi VII$_0$ symmetry is responsible for the fact that such an Ansatz leads to ODEs for the unknown radially dependent functions $D,B,C_{1,2,3}, \phi,A_{1,2}$.

It is clear from the Ansatz above that although there is translational symmetry in the $x_{2,3}$ directions, there is no translational symmetry in the $x_1$ direction. It is replaced by a more complicated one-parameter family of symmetries that involve an $x_1$ translation and at the same time a related rotation in the $x_2-x_3$ plane. This is the reason that this symmetry is also called helical symmetry.

Using this Ansatz we can derive the equations for the various radially-dependent functions and they are spelled out in appendix \ref{app:eoms}.
In section \ref{section:conductivity}, we derive a general formula for the DC conductivity which takes the form \eqref{1}  for all possible black hole within the radial Ansatz \eqref{3}.

Typically we expect to find domain-wall solutions to the equations of motion, which are asymptotically AdS$_5$ in the UV. We will analyse the IR scaling solutions as, in the spirit of \cite{cgkkm} they will determine the possible IR end points of the RG  flows within our metric Ansatz.\footnote{As seen in \cite{ihqcd,Charmousis:2009xr}, extremal backgrounds with metric elements displaying both power and exponential behaviour are only possible in cases where the action parameters $\delta$, $\gamma_{1,2}$ are tuned to specific values.} This is done in section \ref{section:HelicalAn} and \ref{section:HelicalIs}. In section \ref{section:HelicalAn}, we present saddle points which display spatial anisotropy in the metric, and then compute the low frequency, zero temperature asymptotics of the AC conductivity, as well as the low temperature asymptotics of the DC conductivity. In section \ref{section:HelicalIs}, we study saddle points with irrelevant translation-breaking deformations, as well as their DC conductivity. We also comment on their semi-locally critical limits. Appendix \ref{app:NEC} contains constraints from the Null Energy Condition. Many technical details about the various solutions have been relegated to appendix \ref{app:GroundStatesIs}-\ref{section:ExactTransInv}, while the fluctuation equations for the conductivity can be found in appendix \ref{app:conductivityFluct}.

\subsection{Summary and Outlook \label{Summary}}

In this work, we obtain broad families of extremal backgrounds with helical symmetry, and characterize them by their behaviour under rigid scaling transformations \eqref{ScalingBianchi}. We are interested in saddle points where the effects of translation symmetry breaking are strong, imprinting some anisotropy between the helix director and the transverse plane, or even between the $x_2$, $x_3$ directions of the transverse plane; and in saddle points where translation symmetry breaking is mediated by irrelevant deformations.

We also compute their conductivity, in the AC regime at zero temperature using a matched asymptotics argument, and in the DC regime through a general formula evaluated at the event horizon, relying on the existence of a radial constraint. This gives a formula for the DC conductivity, \eqref{DC42}, made up of a quantum critical term and a dissipative term as in \eqref{1}.

The anisotropic backgrounds are captured at leading order by the Ansatz
\be\label{AnisoIntro}
\begin{split}
&\ud s^2=r^{2\theta/3}\left[-\frac{\ud t^2}{r^{2z_1}}+\frac{L^2\ud r^2+\omega_1^2}{r^2}+\frac1{r^{2z_2}}\left(\omega_2^2+\frac{\lambda}{k^2r^2}\omega_3^2\right)\right]\left(1+O(k^2r^2)\right),\\
&\phi=\kappa\ln r+O(k^2r^2)\,,\quad A_1=Q_1 r^{\zeta-z_1}\left(1+O(k^2r^2)\right)\ud t\,,\quad A_2=Q_2 \left(1+O(k^2r^2)\right)\omega_2
\end{split}
\ee
and behave under rigid scale transformations as
\begin{align}
\label{ScalingBianchi}
& r\to\xi r\,,\quad t\to\xi^{z_1}t\,,\quad x_1\to\xi x_1\,,\quad (x_2,x_3)\to\xi^{z_2}(x_2,x_3)\,,\quad k\to\xi^{-1}k\notag\\
& \phi\to\phi+\kappa\,\ln\xi\,,\quad \ud s^{2}_{k}\rightarrow \xi^{2\theta/3}\,\ud s^{2}_{\xi^{-1}\,k},\quad A_{1}\rightarrow \xi^{\zeta}\,A_{1},\quad A_{2,}{}_{k}\to \xi^{z_{2}}A_{2,}{}_{\xi^{-1}\,k}
\end{align}
$\theta$ and $\zeta$ parameterize the scale covariance of the metric and $A_1$ under \eqref{ScalingBianchi}, and are respectively metric \cite{cgkkm,sachdev} and vector hyperscaling violating exponents \cite{gk2012,g2013}.\footnote{Another exponent related to the scale invariance of the Maxwell term in the action was proposed in \cite{Gath:2012} in the context of EMD theories with a massive vector.} They are also related to the scaling of entropy and DC conductivity with temperature.

Note that \eqref{ScalingBianchi} is not a symmetry of the background, but only of the equations of motion, as suggested by the field transformations in the second line: as such, it generates a new solution related to the original one by a marginal deformation and moreover scaling the $x_1$ coordinate changes the period of the helical lattice $k$ which is meant to be fixed in the UV.

The electric potential can behave in two ways:
\begin{itemize}
\item Either it gives contributions in the stress-tensor appearing at the same power of the radial coordinate as logarithmic derivatives coming from the Einstein tensor. The density has zero scaling dimension and corresponds to a marginal deformation of the solution in the spectrum of static, radial deformations. Then, $\zeta=\theta-2-2z_2$.
\item Or it gives contributions in the stress-tensor appearing at a subleading power of the radial coordinate compared to logarithmic derivatives coming from the Einstein tensor.  The density no longer has zero scaling dimension and it now generates an irrelevant deformation of the background. $\zeta$ is now unfixed, but $z_1=3z_2/2$.
\end{itemize}

The DC scaling of the states \eqref{AnisoIntro} is captured both for marginal/irrelevant density deformation by
\be\label{DCscalingIntro}
\sigma_{DC}\sim T^{\frac{\zeta -2}{z_1}}
\ee
which is of the same form as that seen in \cite{g2014}.\footnote{This anomalous behaviour can be predicted by scaling analysis when temperature is the only scale \cite{g2014,Karch:2014mba}. Note that we can restore a hyperscaling violating scale $\ell$ to render the running scalar dimensionless, $\phi=\kappa\ln(r/\ell)$, as well as to make up for the missing dimensions in the solution. It is a marginal deformation. Upon doing this, the DC conductivity becomes $\sigma_{DC}\sim T^{\frac{\zeta -2}{z_1}}\ell^{\zeta-3}$, which restores it usual dimension, $d-2=1$ in five bulk dimensions.} This suggests that the parameterization of this quantum ciritical contribution in terms of the exponent $\zeta$ might be universal.

It is also of interest to match the scaling behaviour of the DC conductivity with the low-frequency scaling of the AC conductivity at zero temperature. Near (isotropic) quantum critical points, the conductivity is expected to be scale-covariant and behave as $T^{(d-2)/z_1}F(\omega/T)$ with $F(x)\sim x^{(d-2)/z_1}$ for $x\gg1$, $F(x)\sim 1$ for $x\ll1$, \cite{DS}. The scaling in temperature/frequency of the DC conductivity/zero temperature, low frequency conductivity should be identical and scale like $T^{(d-2)/z_1}$ or $\omega^{(d-2)/z_1}$, respectively. The former differs from \eqref{DCscalingIntro}, unless $\zeta=2+2z_2-\theta$,\footnote{Accounting for spatial anisotropy and hyperscaling violation.} which is forbidden by our parameter space.

When the density deformation is marginal, we obtain only insulating states, where the optical conductivity behaves covariantly at zero temperature and low frequencies as $\sigma(T\ll\omega\ll\mu)\sim\omega^{(\zeta-2)/z_1}$. Both terms in the DC conductivity have the same temperature dependence, but can be parametrically separated by the ratio of the charge density over $k$.

When the density deformation is irrelevant, the DC conductivity leading temperature dependence comes from the quantum critical contribution. We obtain either insulators with a scale covariant-conductivity, or metals where the conductivity is not always scale-covariant, as summarized in figure \ref{Fig1}.\footnote{This mismatch was also noted in previous works involving spatially dependent scalars, \cite{Donos:2014uba,g2014}.} Moreover, the AC scaling with frequency can vanish, suggesting that a delta function (unrelated to translation symmetry) might be present at zero frequency.

For $\zeta=2$, both the AC and DC conductivities are constant in the IR regime, while for $\zeta=2-3z_2$, only the AC conductivity is so, while the system conducts. This is perhaps reminiscent of the transition between metals and insulators in under-doped cuprates, \cite{Uchida:1991,Uchida:1996}.

The limit where $z=z_1/z_2$, $\tilde\theta=\theta/z_2$, $\tilde\zeta=\zeta/z_2$ are kept fixed while $z_{1,2},\theta,\zeta\to+\infty$ can be taken in \eqref{AnisoIntro}, leading to an exact solution of the form
\be\label{SLIntro}
\begin{split}
&\ud s^2=r^{\frac23\tilde\theta}\left[-\frac{\ud t^2}{r^{2 z}}+\frac{L^2 dr^2+\omega_2^2+\lambda\omega_3^2}{r^2}+\omega_1^2\right],\\
& A_1=Q_1r^{\tilde\zeta- z}\ud t,\quad A_2=Q_2 r^{a_2}\omega_2\,,\quad \phi=\kappa\ln r\,.
\end{split}
\ee
Under rigid scale transformations, the $x_1$ direction does not scale:
\bea
\label{ScalingPartiallyHV}
 &r\to\xi r\,,\quad t\to\xi^{z}t\,,\quad (x_2,x_3)\to\xi(x_2,x_3)\,,\notag\\
& \phi\to\phi+\kappa\,\ln\xi\,,\qquad \ud s^{2}_{k}\rightarrow \xi^{2\tilde\theta/3}\,\ud s^{2}_{k},\quad A_{1}\rightarrow \xi^{\tilde\zeta}\,A_{1},\quad A_{2}\to \xi^{a_{2}+1}A_{2\,,k}
\eea
prompting us to interpret them as partially hyperscaling violating \cite{kasa}.

An analysis of their DC conductivity shows these states are metallic with a power-law scaling
\be
\sigma_{DC}\sim T^{\frac{\tilde\zeta}{z}}
\ee
which can also be obtained from \eqref{DCscalingIntro} by the appropriate limit. For a marginal density deformation $\tilde\zeta=\tilde\theta-2$, the resistivity is found to scale with temperature like the entropy density, which is very reminescent of the mechanism of linear resisitivity found in \cite{DSZ} for semi-locally critical states in massive gravity.

Backgrounds where the metric does not break translation invariance at leading order are just the usual hyperscaling violating backgrounds
\be\label{IsoIntro}
\ud s^2=r^{2\theta/3}\left[-\frac{\ud t^2}{r^{2 z_1}}+\frac{L^2\ud r^2+\ud x_1{}^2+\ud x_2{}^2+\ud x_3{}^2}{r^2}\right],\quad A_1=Q_1r^{\zeta- z_1}\ud t\,,\quad \phi=\kappa\ln r\,,
\ee
which behave under rigid scale transformations as:
\bea
\label{ScalingHV}
 &r\to\xi r\,,\qquad t\to\xi^{\tilde z_1}t\,,\qquad (x_1,x_2,x_3)\to\xi(x_1,x_2,x_3)\,,\notag\\
& \phi\to\phi+\kappa\,\ln\xi\,,\qquad \ud s^{2}_{k}\rightarrow \xi^{2\theta/3}\,\ud s^{2}_{k},\qquad A_{1}\rightarrow \xi^{\zeta}\,A_{1}\,.
\eea
They have a semi-locally critical limit where $\theta,z_1,\zeta\to+\infty$ while the appropriate ratios are kept finite, under which the spatial directions no longer scale under rigid scale transformations.

For $z_1<+\infty$, the translation-breaking (lattice) modes are exponentially suppressed, which in turn lead to metallic states with an exponentially suppressed resistivity at low temperatures, rather than with a power law. This is similar to  \cite{semilocal} and in agreement with the expectation that no finite momentum mode can resonate at finite $z_1$, \cite{Hartnoll:2012rj}. In this sense, our results complement those of \cite{semilocal} where frequency-dependent spectral densities were computed.
As far as we know this is the first instance in holography where exponential conductivities are explicitly computed.

This can be evaded in the semi-locally critical limit where $z_1$ diverges and indeed in this case we recover metallic states with  a power-law DC conductivity.

In this work, we have not constructed any finite temperature black hole solutions, whose zero temperature limit would be described by the extremal backgrounds described above. This is clearly an important step which we leave for future work.

We have found that the (boundary) DC conductivity is the sum of a dissipative and quantum critical term. However, it is computed from horizon data, which complicates the boundary interpretation of these two contributions, in particular the quantum critical one. It would be nice to understand better its origin from the boundary point of view.

We have seen that metallic states could be obtained, with a different scaling for the DC conductivity and the zero temperature AC conductivity, even possibly with a zero frequency delta function. Moreover, the DC conductivity can be dominated by the quantum critical contribution, either when it sets the leading low temperature dependence, or if both terms have the same temperature dependence, when translation-breaking effects are strong. Clearly, we would like to understand how this impacts the full frequency-dependent and finite temperature conductivity, in particular whether it displays a sharp Drude-like peak or not.

Finally, we turn to the question of metal/insulator transitions. They can be triggered along two scenarii, either if the metallic phase displays a relevant mode, or if a mode becomes relevant by tuning UV data such as the lattice scale. The same happens for the anisotropic saddle points \eqref{AnisoIntro} which can also display an RG-relevant mode, as shown in figure \ref{fig:ParSpace}. In the semi-locally critical limit \eqref{SLIntro}, the scaling dimension of the mode depends explicitly on $k$, and there is a region of instability when an irrelevant mode sourced by the magnetic field becomes relevant at small $k$, see figure \ref{Fig:k-instability}. The endpoint of this instability can then also be an insulating phase. Concrete UV completions in which such transitions can occur are left for future investigation.

A method to derive the DC heat conductivity has been outlined in \cite{Donos:2014cya}. It would be interesting to adapt it to our setup, and to work out whether the phases presented here can conduct heat or not.

\section{The DC conductivity along the helix director \label{section:conductivity}}

We will now calculate the electric DC conductivity on black hole backgrounds which are captured by the radial Ansatz \eqref{3}. To proceed, we will follow the simple argument developed in \cite{Donos:2014uba} by introducing a constant electric field perturbation on the asymptotic boundary and reading off the response of the current.

More concretely, we will assume that our background geometry \eqref{3} asymptotes to $AdS_{5}$ as $r\to\infty$ and that there is a regular Killing horizon at $r=r_{+}$. Without loss of generality, we will assume that $Z_{i}\left(\phi(\infty) \right)\rightarrow 1$ and $V\left(\phi(\infty) \right)\rightarrow 12$ in the action \eqref{Action3}. Following the logic of \cite{Donos:2014uba} we consider the perturbation
\begin{align}
\delta\,ds^{2}=C_{1}\,\left(\delta g_{t1}\,dt+\delta g_{r1}\,dr\right)\,\omega_{1}+C_{3}\,\delta g_{23}\,\omega_{2}\,\omega_{3}\notag\\
\delta A_{1}=\left(-E\,t+\delta a_{1}\right)\,\omega_{1},\quad \delta {A}_{2}=\delta b\,\omega_{3}
\end{align}
where all the functions $\left\{\delta g_{t1},\, \delta g_{r1},\delta g_{23},\, \delta a_{1},\, \delta b\right\}$ used to parametrize the perturbation depend only on $r$ while $E$ is the amplitude of the constant boundary electric field.

Without loss of generality, we will fix a gauge for our background \eqref{3} in which $B^{-1}=D=U$. In this gauge, a regular horizon at $r=r_{+}$ will yield the leading order behaviour $U\approx 4\pi T\left(r-r_{+}\right)+\cdots$ , $A_{1} \approx A_{1}^{+}\,\left(r-r_{+}\right)+\cdots$ with $T$ being the Hawking temperature and all other functions in \eqref{3} taking constant values. We wish to impose ingoing boundary conditions on the horizon and this can be achieved by requiring\footnote{It is only then that our perturbation will only depend on the regular ingoing Eddington-Finklestein coordinates $r$ and $v=t+\frac{1}{4\pi T}\,\ln\left(r-r_{+}\right)$. In particular, we will have that $\delta A_{1}\approx \,-E\,v\,\omega_{1}$ and $\delta g_{t1}\,dt+\delta g_{r1}\,dr\approx dv$.}
\begin{align}\label{eq:DC_nh_regcondition}
\delta a_{1}^{\prime}=-\frac{E}{U}+\cdots,\quad \delta g_{t1}=U\,\delta g_{r1}+\cdots
\end{align}
at $r=r_{+}$ and with $\delta g_{t1}$ taking a constant value i.e. $U\,\delta g_{r1}$ admits a Taylor expansion on the horizon. In addition, we require that the functions $\left\{ \delta g_{23},\,\delta b\right\}$ are analytic at $r=r_{+}$. On the $AdS_{5}$ boundary we will require that all the functions $\left\{\delta g_{t1},\, \delta g_{r1},\delta g_{23},\, \delta a_{1},\, \delta b\right\}$ will have normalizable fall offs. At this point we have to stress that the functions $\delta a_{1}$ and $\delta g_{t1}$ satisfy second order equations which allow for boundary conditions of the form \eqref{eq:DC_nh_regcondition}. For the function $\delta a_{1}$ this can be seen from the gauge field equation of motion with which we are dealing next.

The equations of motion of the gauge field $A_{1}$ can be written in an integrated form
\begin{align}\label{eq:bulk_current}
J=\sqrt{C_{1}C_{2}C_{3}}\,\left(-C_{1}^{-1}\,U\,Z_{1}\,\delta a_{1}^{\prime}-\,Z_{1}\,A_{1}^{\prime}\,\delta g_{t1}\right)\sp J'=0\,.
\end{align}
On the $AdS_{5}$ boundary at $r\rightarrow \infty$ we have that $U\approx C_{i}\rightarrow r^{2}$. Moreover, for the current perturbation we have that $\delta a_{1}\propto -\frac{j}{r^{2}}$ with $j$ being the current we are after. Since we require a normalizable fall off for $\delta g_{t1}\propto \mathcal{O}(r^{-1})$ it follows that $J$ is precisely the current we would like to calculate. On the other hand, we can evaluate this expression on the black hole horizon at $r=r_{+}$. The first term can be easily evaluated by using the first boundary condition in \eqref{eq:DC_nh_regcondition}. For the second term we will have to take a closer look at the linearized equations of motion. Doing so reveals that we can algebraically solve for the function $\delta g_{r1}$
\begin{align}
L^{2}\,\delta g_{r1}=&-U^{-1}\frac{E}{k}\,C_{2}\,C_{3}\,Z_{1}\,A_{1}^{\prime}+C_{3}\,\left(C_{3}-C_{2} \right)\,\delta g_{23}^{\prime}\nn\\
&+Z_{2}\,\left( -C_{2}\,A_{2}\,\delta b^{\prime}+C_{3}\,A_{2}^{\prime}\,\delta b\right) + C_{3}\,\left( 2\,C_{2}\,\left( \frac{C_{2}^{\prime}}{C_{2}}-\frac{C_{3}^{\prime}}{C_{3}}\right) +Z_{2}\,A_{2}\,A_{2}^{\prime}\right)\,\delta g_{23} \nn\\
L^{2}=&k\,\left(C_{2}-C_{3} \right)^{2}+k\,C_{2}\,Z_{2}\,A_{2}^{2}\,.
\end{align}
From the above equation and in combination with the analyticity properties, we deduce that close to the horizon
\begin{align}
\delta g_{r1}=-U^{-1}\frac{E}{k^2}\,\frac{C_{2}\,C_{3}\,Z_{1}\,A_{1}^{\prime}}{\left(C_{2}-C_{3}\right)^{2}+C_{2}Z_{2}\,A_{2}^{2}}+\mathcal{O}\left(\left(r-r_+\right)^0 \right).
\end{align}
Using the second boundary condition in \eqref{eq:DC_nh_regcondition} we find that on the horizon
\begin{align}
\delta g_{t1}=-\frac{E}{k^2}\,\frac{C_{2}\,C_{3}\,Z_{1}\,A_{1}^{\prime}}{\left(C_{2}-C_{3}\right)^{2}+C_{2}Z_{2}\,A_{2}^{2}}+\mathcal{O}\left(r-r_{+}\right).
\end{align}

The above allows us to express the DC conductivity in terms of the black hole horizon data as
\begin{align}\label{DC42}
\sigma_{DC}&=\sqrt{\frac{C_{2}C_{3}}{C_{1}}}\,\left( Z_{1}+\frac{1}{k^{2}}\,\frac{Q^{2}}{\left(C_{2}-C_{3} \right)^{2}+C_{2}\,Z_{2}\,A_{2}^{2} }\right)\left.\right|_{r=r_{+}}\\
Q&=-Z_{1}\,\sqrt{C_{1}C_{2}C_{3}}\,A_{1}^{\prime}\,.
\end{align}
with $Q$ being the charge density.

In the following sections, we will discuss the nature of charge transport mediated by saddle points with helical symmetry. We will use the formula for the DC conductivity \eqref{DC42} in order to derive the leading scaling behaviour in temperature.

Though we have not constructed families of black holes whose extremal limit reduces to one of the ground states we will shortly describe, we have shown that a small temperature deformation could always be turned on consistently. Since we are only interested in the leading order behaviour of the DC conductivity, we can simply evaluate \eqref{DC42} at the horizon radius $r_+$ using the metrics of section \ref{section:HelicalAn} and \ref{section:HelicalIs}, and then use the fact that $r_+^{-z_1}\sim T$. This is entirely similar to the manipulations involved in deriving the low-$T$ scaling of the thermal entropy.

\section{Anisotropic metals and insulators with helical symmetry\label{section:HelicalAn}}

In the next subsection \ref{subsection:HelicalAn}, we present in detail anisotropic IR solutions with helical symmetry with either marginal or irrelevant density deformation. We also present the partially hyperscaling violating limit where the $x_1$ direction decouples from the scaling \eqref{ScalingBianchi}. Then, we turn to the low-frequency, zero temperature asymptotics of the AC conductivity in section \ref{subsection:HelicalAnAC}. Finally, we examine the low-temperature scaling of the DC conductivity in section \ref{subsection:HelicalAnDC}.

\subsection{Anisotropic IR saddle points\label{subsection:HelicalAn}}

Recalling the metric Ansatz \eqref{3}, let us assume $C_2(r)\neq C_3(r)$. At leading order in the radial coordinate, this can be realised either when these two functions scale differently with $r$, or if they have a different overall prefactor. We focus on each possibility in turn.

From the equation of motion \eqref{E32}, anisotropic solutions with different $r$ dependence for $C_2$ and $C_3$ cannot be expressed in closed form when $k\neq0$, but rather as power series expansions controlled by powers of $r$. To connect with previous studies of IR fixed points in EMD theories \cite{cgkkm, gk2012,g2013,g2014}, note that here it is not a stress-tensor contribution from the matter fields which gives a subleading term in the IR, but rather a term coming from the metric via the Einstein tensor which does not take the form of a logarithmic derivative. The solutions are captured by the following scaling Ansatz
\be\label{sol422}
\begin{split}
&\ud s^2=-D(r)\ud t^2+B(r)\ud r^2+C_1(r)\omega_1^2+C_2(r)\omega_2^2+C_3(r)\omega_3^2\\
&B(r)=L^2r^{\frac23\theta-2}\left(1+\sum_{i=1}b_i \left(kr\right)^{2i}\right), \quad D(r)=r^{\frac23\theta-2z_1}\left(1+\sum_{i=1}d_i\left(kr\right)^{2i}\right),\\
&C_1(r)=r^{\frac23\theta-2}\left(1+\sum_{i=1}c_{(1)i}\left(kr\right)^{2i}\right), \quad C_2(r)=r^{\frac23\theta-2z_2}\left(1+\sum_{i=1}c_{(2)i}\left(kr\right)^{2i}\right),\\
&C_3(r)= r^{\frac23\theta-2z_2}\left(\frac{V_0 z_2}{ \left(1-\theta +z_1+2 z_2\right)(kr)^2}+\sum_{i=0}c_{(3)i}\left(kr\right)^{2i}\right)\\
&\phi=\kappa\log r+\sum_{i=1}\varphi_{i} \left(kr\right)^{2i}, \quad A_2(r)=Q_2\left(1+\sum_{i=2}a_{(2)i} \left(kr\right)^{2i}\right),\\
&L^2= \frac{\left(\theta-2 -z_1-2 z_2\right) \left(\theta-1 -z_1-2 z_2\right)}{V_0}\,,\quad Q_2^2=\frac{2 \left(z_2-1\right)V_0^2 z_2}{ \left(1-\theta +z_1+2 z_2\right)^2k^4}\,.
\end{split}
\ee
In this coordinate system, the expansion is in even powers of $r$, and the various coefficients $a_{(2),i}$, $b_i$, $c_{(1,2,3),j}$, $d_i$ and $\varphi_i$ are completely fixed by the choice of radial coordinate and the equations of motion. All amplitudes are proportional to $k^2$, the periodicity of the helix, which emphasizes the fact that the spatial anisotropy of the geometry is generated by the choice of spatial Bianchi VII$_0$ symmetry. Moreover, consistency of the expansion implies that the IR is $r\to0$.  The geometries always obey the NEC \eqref{NEC2} at leading order in $r$. The magnetic field $A_2$ is constant in the IR at leading order.

 The scaling exponents are related to the parameters of the action:
\be
\kappa^2=\frac{2}{3} \left(\theta ^2-6-6 z_2-6 z_2^2+z_1 \left(6-3 \theta +6 z_2\right)\right),\quad \kappa\delta = \frac{2 \theta }{3},\quad \kappa\gamma _2= \frac{2}{3} \left(\theta-6 -3 z_2\right).
\ee

The fact that a series in powers of $kr$ can be developed in this anisotropic case is in sharp contrast with finite momentum deformations of translation-invariant backgrounds, whose translation-breaking deformations are exponentially suppressed, as we detail in appendix \ref{app:GroundStatesIsThetaZ}. This is in agreement with expectations that their resistivities should be exponentially suppressed: degrees of freedom which obey the nonrelativistic scaling $\omega\sim k^{z_1}$ at low energies cannot live at finite momentum in presence of translation invariance, \cite{Hartnoll:2012rj,semilocal}.

It can be checked that the leading order powers in $r$ of the $tt$ and $\omega_2{}^2$ elements of the metric always vanish in the IR, while the $\omega_{1,3}{}^2$ elements do not necessarily do so. Moreover, the Ricci scalar goes at leading order like $r^{-2\theta/3}$, which means that there is a mild\footnote{In the sense that it can be hidden behind an event horizon.} curvature singularity in the IR given that $\theta>0$ (see the parameter space below in figures \ref{fig:ParSpaceMargIns} and \ref{fig:ParSpace}).

Given the expression for the non-zero temperature deformation (see below), the entropy scales as
\be
S=T^{\frac{2z_2+2-\theta}{z_1}}
\ee
and always vanishes in the allowed parameter space as $T\to0$, ensuring the specific heat is always positive.

 As described in section \ref{Summary}, the density deformation can be marginal or irrelevant. We now detail both cases.

\paragraph{Marginal density deformation\\}
In this case, $A_1$ reads
\be
 A_1(r)=Q_1r^{\theta-2 -z_1-2 z_2}\left(1+\sum_{i=1}a_{(1)i}\left(kr\right)^{2i}\right),
\ee
with
\be
\kappa\gamma _1=4-\frac{4 \theta }{3}+4 z_2\,,\qquad  Q_1^2=\frac{2 z_1-3 z_2}{2-\theta +z_1+2 z_2}
\ee
Taking the constant scalar limit $\kappa=\delta=\gamma_1=\gamma_2=0$, we recover AdS$_2\times\mathbf R^3$ with both an electric and a magnetic field turned on.
Consistency of the solution requires
\be
L^2>0\,,\quad Q_1^2>0\,,\quad Q_2>0\,,\quad\kappa^2>0\,,\quad \frac{V_0 z_2}{1-\theta +z_1+2 z_2}>0
\ee
to ensure that all parameters are real and that the signature of the metric is Lorentzian.

Next, we turn to radial, static deformations around the solution,
\be
\begin{split}
\frac{\Delta B}B=\sum_i c_i^B r^{\beta_i}\,,\quad \frac{\Delta D}D=\sum_i c_i^D r^{\beta_i}\,,\quad \frac{\Delta C_j}{C_j}=\sum_i c_i^j r^{\beta_i}\,,\quad j=1,2,3\\
\frac{\Delta \Phi}\Phi=\sum_i c_i^\phi r^{\beta_i}\,,\quad \frac{\Delta A_2}{A_2}=r^{2}\sum_i c_i^m r^{\beta_i}\,,\quad \frac{\Delta A_0}{A_0}=\sum_i c_i^e r^{\beta_i}\,.
\end{split}
\ee
The following modes can be found:
\be
\beta_{i,\pm}=\frac12(2-\theta +z_1+2 z_2)\pm\nu_i\,.
\ee
We give explicit values to $\nu_i$ to only for two of them (the three other pairs are still of the form above but are given by a 6th order polynomial which does not factorise and yields cumbersome expressions)
\be
\begin{split}
&\nu_1=\frac12(2-\theta +z_1+2 z_2)\,,\\
&\nu_2=3-\frac{\theta }{2}+\frac{z_1}{2}+z_2\,.
\end{split}
\ee
$\beta_{1,\pm}$ are both doubly-degenerate and correspond respectively to a finite temperature and its conjugate marginal mode.
So we require $\beta_{1,+}<0$ given that the IR is $r\to0$. We now have to check that $\beta_{i,+}>0$ with $i=2\dots5$, so they correspond to irrelevant modes. For lack of nice analytical expressions for the modes, it is hard to do in full generality except for $\beta_{2,+}$. Once this is imposed though, a few explicit values seem to show that the other $+$ modes are positive, at least in some region of the parameter space, which reads:
\be\label{ParSpaceRel}
\begin{split}
&13-3 \sqrt{21}<z_2<0\,,\qquad\frac{1}{2} \left(1+2 z_2\right)-\frac{1}{2} \sqrt{3} \sqrt{7+8 z_2}<z_1<\frac{3 z_2}{2}\,,\\
&\frac{3 z_1}{2}+\frac{1}{2} \sqrt{3} \sqrt{8-8 z_1+3 z_1^2+8 z_2-8 z_1 z_2+8 z_2^2}<\theta <4+z_1+2 z_2\,.
\end{split}
\ee
Note that $z_{1,2}<0$, while $\theta>0$. We plot it in figure \ref{fig:ParSpaceMargIns} for $z_2=(-1/10,-1/2)$.

\FIGURE{
\begin{tabular}{cc}
\includegraphics[width=.45\textwidth]{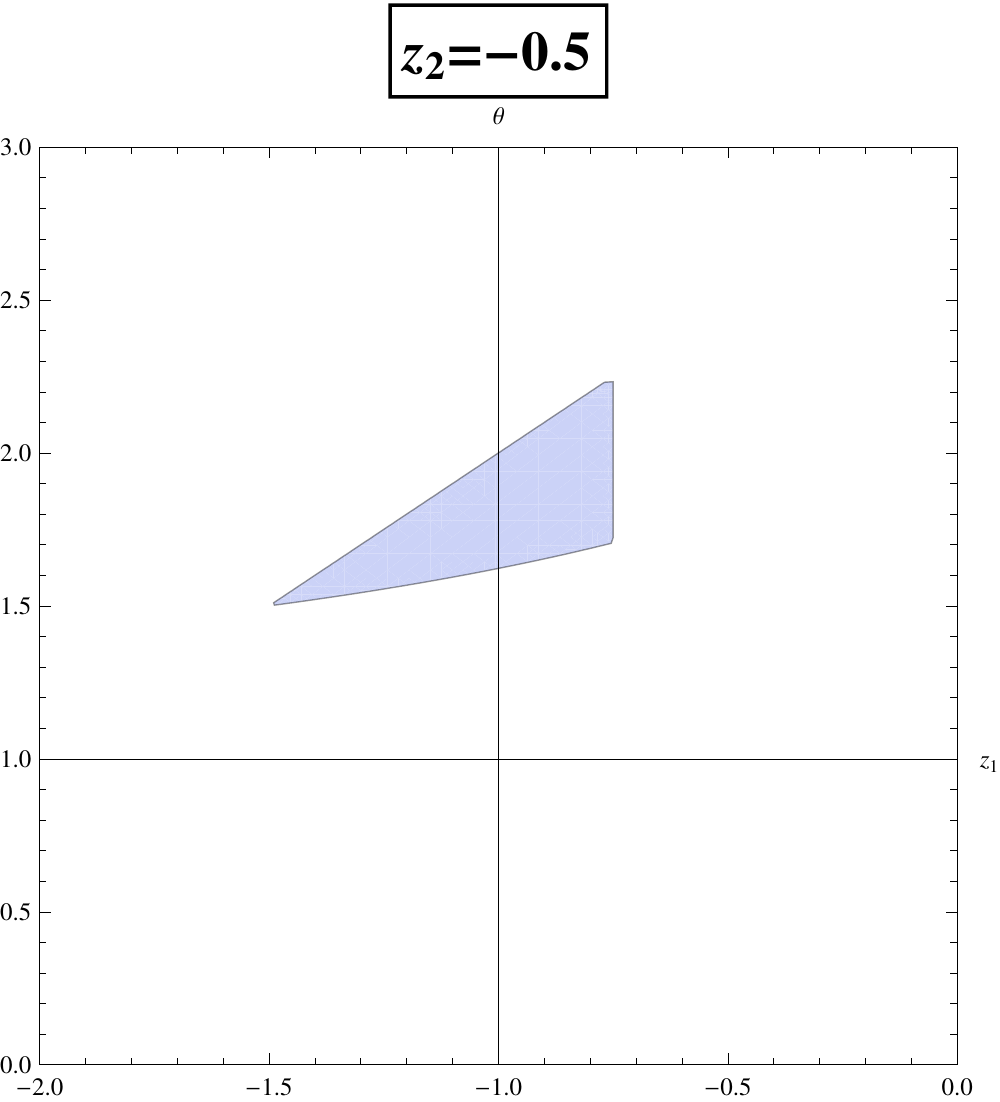}&\includegraphics[width=.45\textwidth]{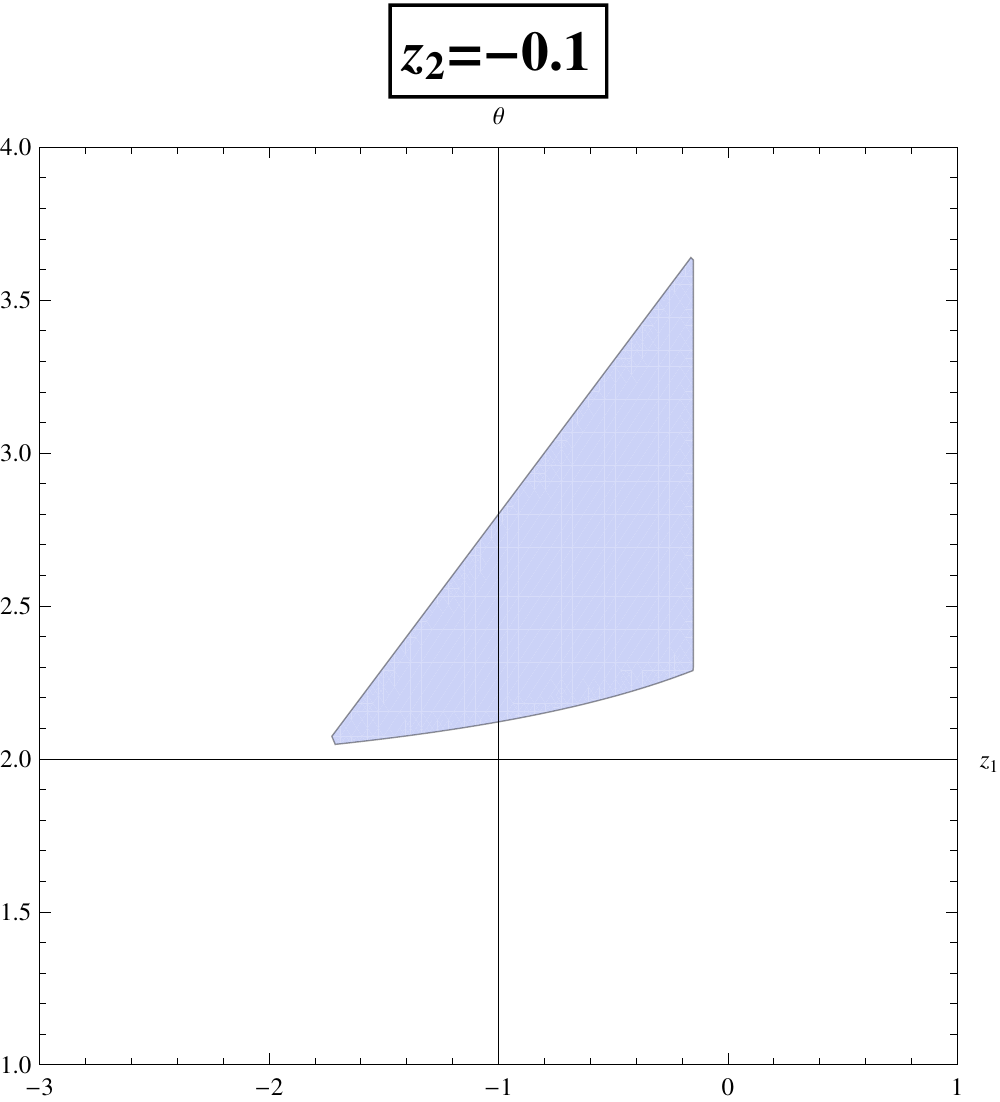}
\end{tabular}
\caption{The parameter space of the ground states \protect\eqref{sol422} with marginal density deformations, for $z_2=-1/2$ (left) and $z_2=-1/10$ (right).}
\label{fig:ParSpaceMargIns}}

\paragraph{Irrelevant density deformation\\}

In this case, $A_1$ enters in the solution as an irrelevant mode, and $z_1=3z_2/2$ is fixed in terms of $z_2$.

Let us parameterize the deformations in the following way
\be
\begin{split}
\frac{\Delta B}B=\sum_{i=1}^4 c_{i,\pm}^B r^{\beta_{i,\pm}}\,,\quad \frac{\Delta D}D=\sum_{i=1}^4 c_{i,\pm}^D r^{\beta_{i,\pm}}\,,\quad \frac{\Delta C_j}{C_j}=\sum_{i=1}^4 c_{i,\pm}^j r^{\beta_{i,\pm}}\,,\quad j=1,2,3\\
\frac{\Delta \Phi}\Phi=\sum_{i=1}^4 c_{i,\pm}^\phi r^{\beta_{i,\pm}}\,,\quad \frac{\Delta A_2}{A_2}=r^{2}\sum_{i=1}^4 c_{i,\pm}^m r^{\beta_{i,\pm}}\,,\quad A_1=r^{\frac{\zeta }{2}+\frac{\theta }{2}-1-\frac52z_2} c_{5,\pm}^e r^{\beta_{5,\pm}}.
\end{split}
\ee
The $A_1$ modes are decoupled from the others at linear order, and are such that
\be
A_1=c_{5,+}^e r^{\zeta-\frac32z_2}+c_{5,-}\,,\qquad \kappa  \gamma _1= 2-\zeta -\frac{\theta }{3}+2 z_2\,.
\ee
We have introduced a conduction exponent $\zeta$ to characterize the scaling of the $+$ mode, through the violation of the helical scaling \eqref{ScalingBianchi}. As we will see later, $\zeta$ also plays a role in the scaling of the conductivity, hence its name. In the previous family of solutions with $A_1$ marginal, it took the value $\zeta=\theta-2-2z_2$.

Using the standard technology, we find the following values for the radial, static perturbations, which we write:
\be
\beta_{i,\pm}=1-\frac{\theta }{2}+\frac{7 z_2}{4}\pm\nu_i
\ee
with
\be
\begin{split}
&\nu_1=1-\frac{\theta }{2}+\frac{7 z_2}{4}\,,\qquad \nu_2=\frac{1}{4} \sqrt{4 (\theta-2 )^2-76 (\theta-2 ) z_2+217 z_2^2}\,,\\
&\nu_3=\frac{1}{4 \left(2 \left(\theta ^2-6\right)+(6-9 \theta ) z_2+6 z_2^2\right)}\left(\surd (16 (\theta-2 )^2 (132-96 \theta -28 \theta ^2+16 \theta ^3+\theta ^4)\right.\\
&-32 \left(-996+546 \theta +614 \theta ^2-447 \theta ^3+32 \theta ^4+16 \theta ^5\right) z_2\\
&+24 (201 \theta ^4-82+2992 \theta -2002 \theta ^2-144 \theta ^3) z_2^2-8 (8838-7089 \theta -3908 \theta ^2+2506 \theta ^3) z_2^3\\
&\left.\left.+\left(-11244-61260 \theta +39769 \theta ^2\right) z_2^4-84 (-386+419 \theta ) z_2^5+11172 z_2^6\right)\right),\\
&\nu_4=3-\frac{\theta }{2}+\frac{7 z_2}{4}\,,\qquad \nu_5=\frac{\zeta }{2}-\frac{3 z_2}{4}\,.
\end{split}
\ee
$\beta_{1,\pm}$ are both doubly-degenerate and correspond respectively to a (relevant) finite temperature and its conjugate marginal mode, which implies $\beta_{1,+}<0$. We now have to check that $\beta_{i,+}>0$ with $i=2\dots5$ in order to correspond to irrelevant modes. This is always true for $\beta_{2,+}$ and $\beta_{3,+}$. The others give some constraints. When either $\beta_{4,+}$ or $\beta_{5,+}$ are relevant, then the system is expected to be RG-unstable, and can mediate a quantum phase transition between an RG-stable metallic and insulating phase, as shown  in \cite{Donos:2012js}.

The constant scalar limit is $\theta=0$, $z_2=-2$ and $\zeta=-2$, which is outside of the region where all the deformations are irrelevant: indeed, $\beta_{4,+}$ and $\beta_{5,+}$ are both relevant in this region. This is in contrast with \cite{Donos:2012js} where these deformations could be irrelevant through the action of a Chern-Simons coupling. Upon taking that limit, we find agreement  with \cite{Donos:2012js}  for the expression of the extremal solution \eqref{sol422}.

The detailed parameter space $(z_2,\theta,\zeta)$ is complicated:
\be\label{ParSpaceIrr}
\begin{split}
&13-3 \sqrt{21}<z_2<0\,,\qquad\frac{9 z_2}{4}+\frac{1}{4} \sqrt{3} \sqrt{32-16 z_2+11 z_2^2}<\theta <\frac{1}{2} \left(8+7 z_2\right),\\
&\zeta >-2+\theta -2 z_2\,,
\end{split}
\ee
 but note that $z_{2}<0$, while $\theta,\zeta>0$. We reproduce it fixing $z_2=-1/2$ or $\zeta=3$ in figure \ref{fig:ParSpace}.

\FIGURE{
\begin{tabular}{cc}
\includegraphics[width=.45\textwidth]{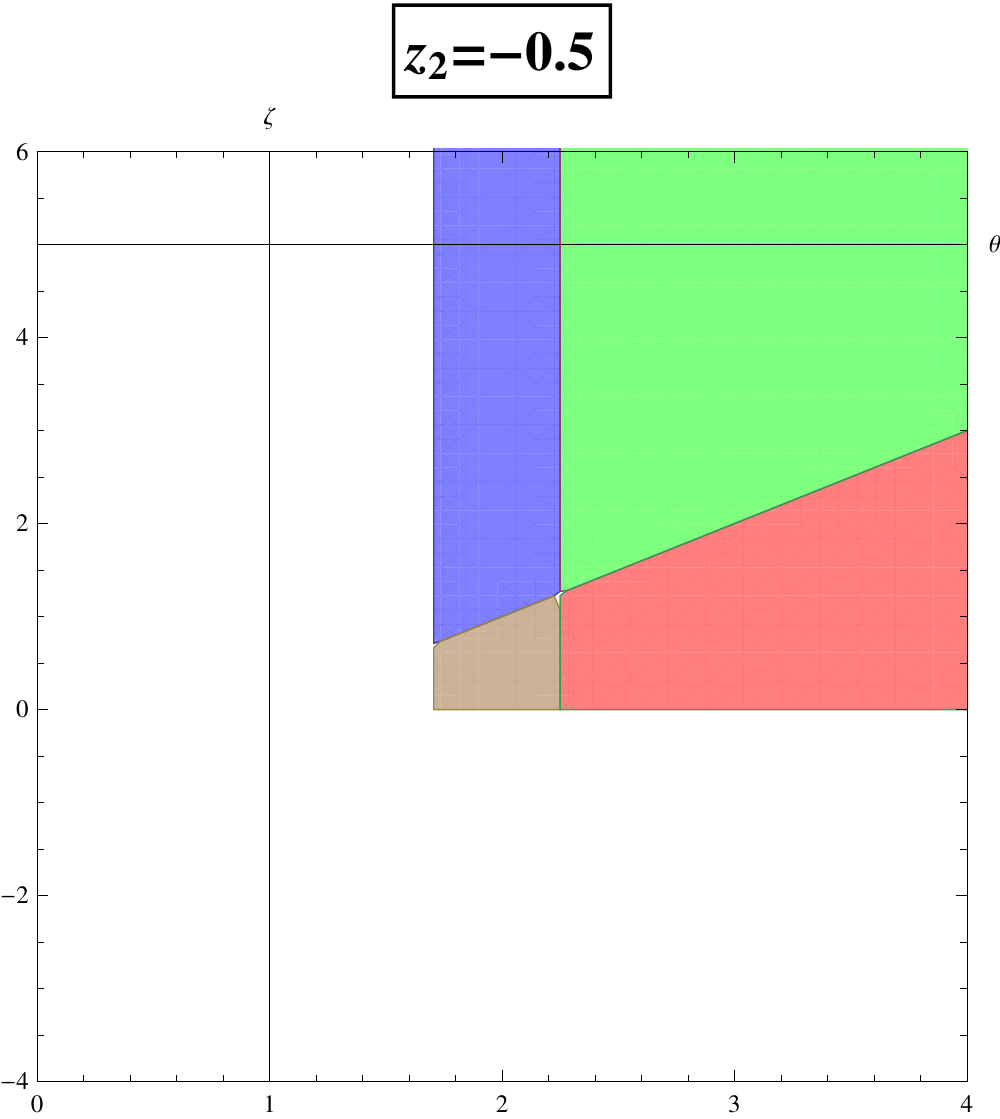}&\includegraphics[width=.45\textwidth]{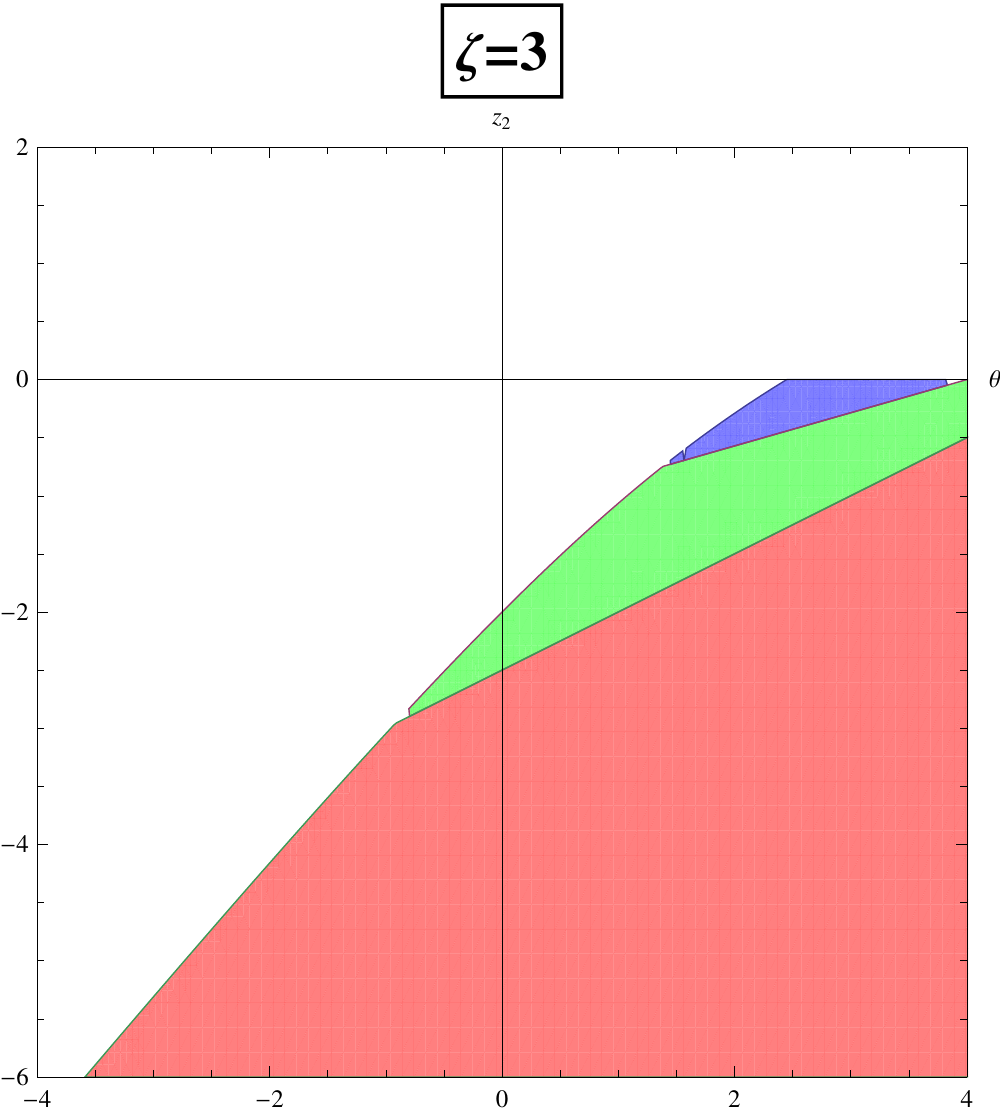}
\end{tabular}
\caption{The parameter space of the ground states \protect\eqref{sol422} with irrelevant electric potential, for $z_2=-1/2$ (left) and $\zeta=3$ (right). In blue, the region where all deformations are irrelevant; in green, where $\beta_{4,+}$ is relevant; in brown, where $\beta_{5,+}$ is relevant; and in red, where both $\beta_{4,+}$ and $\beta_{5,+}$ are relevant.}
\label{fig:ParSpace}}

\paragraph{Partially hyperscaling violating limit $z_2\to+\infty$\\}

Upon taking the limit $z_2\to\infty$, $\theta\to+\infty$, $\zeta\to+\infty$ while keeping $\tilde\theta=\theta/z_2$ and $\tilde\zeta=\zeta/z_2$ finite as well as changing radial coordinate to $r\to r^{1/z_2}$, the power expansion collapses and we recover exact extremal backgrounds. Sending $z_1\to+\infty$ while keeping $z=z_1/z_2$ finite, we obtain
\be\label{solPL1}
\ud s^2=r^{\frac23\tilde\theta}\left[-\frac{\ud t^2}{r^{2z}}+\frac{L^2 dr^2+\omega_2{}^2+\lambda\omega_3{}^2}{r^2}+\omega_1{}^2\right],\quad \phi=\kappa\ln r\,.
\ee
Under this limit, the helix director $\partial_{x_1}$ has decoupled from the rigid scaling transformations \eqref{ScalingPartiallyHV}.
These solutions are presented in greater detail in appendix \ref{app:GroundStatesIsPL}. The magnetic potential can now be non-constant,
\be
A_2=Q_2 r^{a_2}\omega_2\,,
\ee
which was not allowed at finite $z_2$: for $a_2\neq0$, the IR of \eqref{sol422} would lie at $r\to+\infty$ and hence the power expansion was not consistent. The electric potential still behaves
\be
A_1=Q_1r^{\tilde\zeta-z}\ud t\,,
\ee
and two cases must be separated, depending on whether the density sources a marginal (and then $\tilde\zeta=\tilde\theta-2$) or irrelevant deformation (and then $\lambda$ and $z$ are related). When the electric potential is set to zero, they are cousins of solutions presented in \cite{kachru-bia} (which studied solutions with no electric field and a mass turned on for $A_2$, both with and without a running scalar).

\subsection{Low frequency behaviour of the AC conductivity at zero temperature\label{subsection:HelicalAnAC}}

We would like to calculate the low-frequency, zero temperature scaling of the AC conductivity for the solutions of the previous section. After we have also calculated their DC conductivity, we will be able to check whether the conductivity is scale covariant, that is whether it displays the same low frequency and low temperature scaling.

We perturb around the background fields with the following Ansatz
\be
 \delta A_1=e^{-i\omega t}b_1(r) \omega_1\sp \delta A_2=e^{-i\omega t}b_2(r) \omega_3
 \ee
 \be
 \delta(ds^2)=e^{-i\omega t}\left[g_1(r)dt\otimes \omega_1+g_2(r)\omega_2\otimes\omega_3\right]
\ee
and obtain five equations for three propagating modes, \eqref{ea1}-\eqref{ea5}. Keeping $\omega\neq0$ will enable us to calculate the low-frequency, zero temperature scaling of the conductivity.

Our strategy is the following. We will decouple the propagating modes in the IR, by neglecting various terms in the fluctuation equations \eqref{ea1}-\eqref{ea5} which are subleading as $r\to0$. We will be able to express the remaining equations as Schr\"odinger equations, by taking certain linear combinations of the original modes and changing to the radial coordinate
\be\label{SchrCoord}
\frac{\ud \rho}{\ud r}=\sqrt{\frac{B(r)}{D(r)}}\,.
\ee
This will let us access the imaginary part of the IR Green's function of these fields by the standard argument, \cite{Horowitz:2009,KT}. This is not quite yet what we are after, as the conductivity is related to the imaginary part of the UV Green's function. Happily, the argument developed in appendix A of \cite{Donos:2012ra} still applies, and allows to relate the IR and UV data at low frequencies via a matched asymptotics expansion. More explicitly,
\be\label{Im}
\sigma\left(\omega,T\right)=\frac1\omega\Im\left[G^{R,UV}_{\mathcal J^x\mathcal J^x}\left(\omega,T\right)\right]\sim\frac1\omega\sum_I d^I \Im\left[\mathcal G^{R,IR}_{\mathcal{O}_I\mathcal{O}_I}\left(\omega,T\right)\right]
\ee
where the index $I$ runs over all the irrelevant operators $\mathcal{O}_I$ coupling to the current $\mathcal J^x$. The least irrelevant IR operator will dominate the low-frequency scaling of the conductivity.

The two families, with marginal or irrelevant density deformation, have to be treated separately.

\paragraph{Marginal density deformation\\}

After some work, the following changes of variables allows to decouple the linearized equations:
\be
\begin{split}
&\sigma(\rho)=b_2(\rho)-\rho ^{\frac{6-2 \theta +6 z_2}{3 z_1}}\frac{\lambda   g_2(\rho)}{2 Q_2},\\
&\Theta(\rho)= 2 \lambda  \rho ^{\frac{-4+\theta -2 z_2}{2 z_1}} b_2(\rho)+\rho ^{-\frac{\theta -6 z_2}{6 z_1}} Q_2 g_2(\rho) \\
&\Sigma (\rho)=\frac{i k \lambda ^2 \rho ^{\frac{4-\theta +2 z_2}{2 z_1}} \omega  B_0 Q_1 \left(\theta +3 z_1-6 z_2\right) \left(3 z_2-2\right) b_1(\rho)}{6 Q_2 \left(z_1-1\right) z_2}+\frac{\rho ^{\frac{\theta-4 -2 z_2}{2 z_1}}3z_1 \left(3 z_2-2 z_1\right) \sigma '(\rho)}{2 \left(z_1-1\right) \left(\theta +3 z_1-6 z_2\right)},\\
&\Phi(\rho)=\rho ^{\frac{4-\theta +2 z_2}{2 z_1}} b_1(\rho)+\frac{9 i \rho ^{\frac{-4+\theta -2 z_2}{2 z_1}} Q_2 z_1 z_2 \sigma '(\rho)}{k \lambda ^2 \omega  B_0 Q_1 \left(\theta +3 z_1-6 z_2\right){}^2}
\end{split}
\ee
in the Schr\"odinger coordinate \eqref{SchrCoord}.
$\Theta(\rho)$, $\Sigma(\rho)$ and $\Phi(\rho)$ obey the equations:
\be
\begin{split}
&\left(-\frac{\theta ^2-2 z_1 \left(\theta -8 z_2\right)-8 (-3+2 \theta ) z_2+28 z_2^2}{4 \rho ^2 z_1^2}-\omega ^2\right) \Theta+\Theta ''=0\\
&\left(-\frac{\left(-4+\theta -2 z_2\right) \left(-4+\theta +2 z_1-2 z_2\right)}{4 \rho ^2 z_1^2}-\omega ^2\right) \Sigma +\Sigma ''=0\\
&\left(\frac{8 z_1^2-6 z_1 \left(\theta -2 z_2\right)+\left(\theta -2 z_2\right){}^2}{4 \rho ^2 z_1^2}-\omega ^2\right) \Phi+\Phi ''=0\,.
\end{split}
\ee
which give the following contributions to the conductivity, after changing notation to $z_{1,2}, \theta$
\be
\begin{split}
&\sigma\sim \omega^{n_i},\quad i=\Sigma,\,\theta,\,\Phi\\
&n _{\Sigma }=-1+\left|\frac{\left(-4+\theta +z_1-2 z_2\right)}{z_1}\right|,\\
&n _{\Phi }=-1+\left|\frac{\left(\theta -3 z_1-2 z_2\right)}{z_1}\right|,\\
&n_\Theta=-1+\sqrt{\frac{\theta ^2+z_1^2-2 z_1 \left(\theta -8 z_2\right)-8 (-3+2 \theta ) z_2+28 z_2^2}{z_1^2}}\,.
\end{split}
\ee
Given the parameter space \eqref{ParSpaceRel}, $\sigma_\Sigma$ is always the most relevant contribution to the optical conductivity, and moreover $n_\Sigma>0$, so the power tail in the frequency always decays as $\omega\to0$.

\paragraph{Irrelevant density deformation\\}

In this case, it is clear that \eqref{ea1} yields an independent equation for the $b_1$ perturbation in the IR limit, which is expected since the electric potential sources an irrelevant deformation. It straightforwardly yields a Schr\"odinger potential equal to
\be
V_{b_1}(\rho)=\frac{(\zeta-2 ) \left(\zeta-2 +3 z_2\right)}{9 z_2^2\rho^2}
\ee
in the Schr\"odinger coordinate \eqref{SchrCoord}.

The other equations \eqref{ea2}-\eqref{ea5} remain coupled for the variables $g_1$, $g_2$ and $b_2$. After substituting in the generic Ansatz for the fields and neglecting the appropriate powers of $r$ by taking the IR limit, they can be decoupled in the following variables
\be
\begin{split}
& \Theta(\rho)=-\frac{3 z \rho ^{-\frac{12+4 z-3 \theta }{6 z}} b_2(\rho)}{z-\theta }-\frac{3 z \rho ^{\frac{2}{3}-\frac{\theta }{6 z}} Q_2 g_2(\rho)}{2 z \lambda -2 \theta  \lambda }\\
&\Sigma(\rho)=\rho^{-\frac{12+4 z-3 \theta }{6 z}}\frac{\ud}{\ud\rho}\left[\rho ^{-\frac{6+4 z-2 \theta }{3 z}} b_2(\rho)-\frac{\lambda  g_2(\rho)}{2 Q_2}\right],
\end{split}
\ee
as well as changing to the Schr\"odinger coordinate \eqref{SchrCoord}. The two modes obey Schr\"odinger equations:
\be
\begin{split}
0=&\frac{d^2\Theta}{d\rho^2}+\left[\omega^2+\frac{\theta ^2+24 z_2-19 \theta  z_2+52 z_2^2}{9 z_2^2\rho^2}\right]\Theta\,,\\
0=&\frac{d^2\Sigma}{d\rho^2}+\left[\omega^2+\frac{\left(5 z_2-\theta\right) \left(8 z_2-\theta\right)}{9 z_2^2\rho^2}\right]\Sigma\,.
\end{split}
\ee

Now that we have all the effective Schr\"odinger potentials, we can use the standard matching argument and solve the Schr\"odinger equations in terms of a Hankel function with ingoing boundary conditions at the horizon. We obtain
\be
\begin{split}
&\sigma\sim \omega^{n_i},\quad i=\Sigma,\,\theta,\,b_1\\
&n _{\Sigma }=-1+\left|\frac{13}3 -\frac{2 \theta }{3z_2}\right|,\\
&n _{b_1}=-1+\left|\frac{-4+2 \zeta +3 z_2}{3z_2}\right|,\\
&n_\theta=-1+\frac{1}{3} \sqrt{\frac{4 \theta ^2+96 z_2-76 \theta  z_2+217 z_2^2}{z_2^2}}\,.
\end{split}
\ee
These exponents reduce to the values found in \cite{Donos:2012js} in the constant scalar limit, $\theta=0$, $z_2=-2$ and $\zeta=-2$. The exponent $n_{b_1}$ is extremely similar to those found in \cite{g2013,g2014} in the case where $k=0$. We may now proceed and determine which is the most relevant in our parameter space. It is straightforward to observe  that the powers for $n_\theta$ and $n_\Sigma$ are always positive, while $n_{b_1}$ can become negative.

We plot this in fig. \ref{Fig1} for $z_2=-1/2$. In the blue and red regions $n_{b_1}$ dominates and is positive, while in the green and brown region it dominates but is negative. In the purple region, $\sigma_\theta$ dominates and is positive.

\subsection{Low temperature behaviour of the DC conductivity\label{subsection:HelicalAnDC}}

We examine in turn the two anisotropic families \eqref{sol422}, distinguishing between marginal and irrelevant density deformation. Remember that in the marginal case, $z_1$ and $z_2$ are unrelated and $\zeta=\theta-2-2z_2$, while $z_1=3z_2/2$ with $\zeta$ unfixed in the irrelevant case. We also contrast the DC scaling with the AC scaling derived previously.

\paragraph{Solution \protect\eqref{sol422} with marginal density deformation\\}

Both the quantum critical and dissipative terms scale identically with temperature in the DC conductivity:
\be
\sigma_{DC}\sim T^{\frac{-4+\theta -2 z_2}{z_1}}\left(1+O(T^{-2/z_1})\right)+\cdots\sim T^{\frac{\zeta-2}{z_1}}\left(1+O(T^{-2/z_1})\right)+\cdots
\ee
where the extra terms are always subleading since $z_1<0$ in the parameter space.

Moreover, $\s_{DC}$ always vanishes at zero temperature, so these solutions are insulators. We found previously that the zero temperature optical conductivity in this case was given by
\be
\sigma\sim\omega ^{\left|\frac{-4+\theta -2 z_2}{z_1}+1\right|-1}\sim\omega ^{\frac{-4+\theta -2 z_2}{z_1}}
\ee
where the parameter space allows to drop the absolute value. Then, the $\omega$ scaling above matches the DC scaling, so the power tail always vanishes at zero frequency, as expected for an insulator: no spectral weight remains at low frequencies as it has been transferred to higher frequencies.

\paragraph{Solution \protect\eqref{sol422} with irrelevant density deformation\\}

The DC conductivity reads  at leading order
\be
\sigma_{DC}\sim T^{\frac{2 (\zeta -2)}{3 z_2}} \sqrt{\lambda }\left(1+O(T^{-2/z_1})\right)+\frac{Q^2 T^{\frac{2 \left(-4+\theta -2 z_2\right)}{3 z_2}} \sqrt{\lambda }}{k^2 \left(Q_2^2+\lambda ^2\right)}\left(1+O(T^{-2/z_1})\right)+\cdots
\ee
 It can be checked that the first, quantum critical term gives the leading temperatude dependence. The state can be metallic, though it is unclear whether it has a sharp Drude peak or not. It can also be an insulator if
\be
-2+\theta -2 z_2\leq\zeta <2
\ee
where the lower bound is saturated when the density deformation becomes marginal.

\FIGURE{\includegraphics{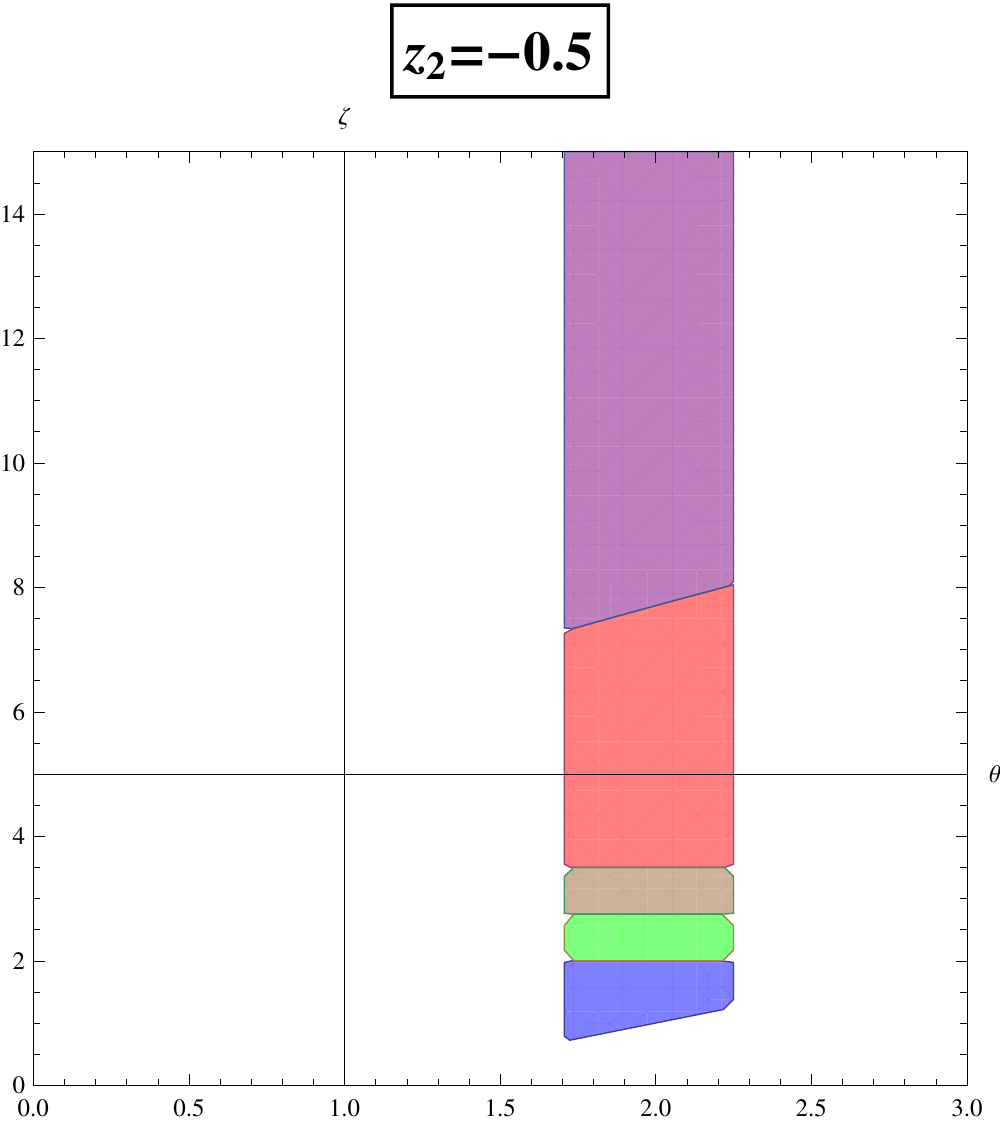}
\caption{The solutions \protect\eqref{sol422} with irrelevant density deformation are insulating in the blue region with a decaying frequency tail given by $\sigma_{b_1}$, $\zeta<2$; metallic with a diverging frequency tail given by $\sigma_{b_1}$ in the green and brown regions $2<\zeta<2-3z_2$; metallic with a decaying frequency tail given by $\sigma_{b_1}$ in the red region; metallic with a decaying frequency tail given by $\sigma_{\theta}$ in the purple region. The scaling with frequency of the power tail agrees with the DC conductivity scaling with temperature in the blue and green regions, for $\zeta<2-3z_2/2$.}
\label{Fig1}}

We found previously that the conductivity could be given either by the two following operators
\be
\begin{split}
\sigma_{b_1}\sim&  \,\omega^{\left|\frac{2\left( \zeta-2 \right)}{3 z_2}+1\right|-1}\,,\\
\sigma_\theta\sim &\,\omega^{\frac{1}{3} \sqrt{\frac{4 \theta ^2+96 z_2-76 \theta  z_2+217 z_2^2}{z_2^2}}-1}\,.
\end{split}
\ee
Only $\sigma_{b_1}$ with a positive absolute value can match the DC scaling, when:
\be
-2+\theta -2 z_2<\zeta <2-\frac32 z_2
\ee
which englobes the whole insulating region, as well as part of the metallic region (in which the optical conductivity then blows up at zero frequency). This is the scale covariant regime, characterized by the conduction exponent being bounded from above and below by other scaling exponents. At the transition to the non scale-covariant regime $\zeta=2-3z_2/2$, the resistivity is linear with temperature, while the AC conductivity has a $1/\omega$ tail.

It can be verified that
\begin{itemize}
\item Whenever the system insulates (blue), the conductivity is given by $\sigma_{b_1}$ and matches the DC scaling with a decaying frequency power tail.
\item When $\sigma_\theta$ dominates (purple), the system is metallic with a decaying frequency power tail. In this case a delta function might be be present at zero frequency to make up for the missing spectral weight, but it would not be connected to translation symmetry and its coefficient unrelated to the density of charge carriers.
\item When the system is metallic and  $\sigma_{b_1}$ dominates, it can have a diverging frequency power tail and match the DC scaling (green), a diverging frequency power tail which does not match the DC scaling (brown), or a decaying frequency power tail (red, which of course does not match the DC scaling).
\end{itemize}

When $\zeta=2$, at the transition between the insulating and metallic regions, both the DC conductivity and the optical conductivity are constant at leading order in temperature and frequency, possibly with log corrections in $\omega$ and $T$.

At the transition between the brown and red region, the AC conductivity exponent changes sign but the DC one does not. For $\zeta=2-3z_2$, we thus have that the AC conductivity is constant plus log corrections, while the DC conductivity scales like $T^{-2}$. This is to be contrasted with the other transition in the scale covariant regime between insulating (blue) and metallic (green) where both the AC and DC are constant with possible log corrections. This is reminiscent of the behaviour seen at the transition between metallic and insulating phases of under-doped cuprates, \cite{Uchida:1991}.

\paragraph{Partially hyperscaling violation solution \protect\eqref{solPL1}\\}
Finally, we come to  the partially hyperscaling violating solutions \eqref{solPL1}, for which the density deformation can also be marginal or irrelevant.
The DC conductivity takes the general form
\be
\sigma_{DC}\sim T^{\frac{\tilde\zeta}{z}}+\frac{Q^2}{k^2} T^{\frac{\tilde\theta-2}{z}}+\cdots
\ee
and always diverges at low temperature: the system is always metallic.

When the density deformation is marginal, $\tilde\zeta=\tilde\theta-2$, so both the quantum critical and dissipative terms scale identically with temperature, but can be distinguished by the value of the ratio $Q^2/k^2$. Intriguingly, the DC scaling is identical to the thermal entropy scaling, $S\sim T^{-\frac{\tilde\theta-2}z}$, so positivity of the specific heat is sufficient to argue that the system is metallic. This is very similar to the results in \cite{DSZ}, which obtained a similar result for semi-locally critical states in massive gravity.

 When the density deformation is irrelevant, the DC conductivity is always dominated by the quantum critical contribution and no longer proportional to the inverse of the entropy density.

\section{Isotropic metals with helical symmetry\label{section:HelicalIs}}

 We briefly describe in this section simpler saddle points, which are exact solutions of the equations of motion and isotropic in the spatial directions at leading order in the metric. We also report on their semi-locally critical limit. In \ref{subsection:HelicalIsDC}, we examine the low-temperature asymptotics of the DC conductivity.

\subsection{Isotropic IR saddle points\label{subsection:HelicalIs}}

In this section, we would like to report the existence of a number of simpler IR saddle points where translation symmetry is not broken at leading order in the metric, but by the static deformations or by the magnetic field. Details can be found in appendix \ref{app:GroundStatesIs}. The simplest option is to start by asking whether hyperscaling violating solutions at finite $z_1$ and $\theta$ \cite{cgkkm} can exist, with Bianchi VII$_0$ deformations:
\be\label{solThetaZ1}
\ud s^2=r^{\frac23\theta}\left[\frac{L^2\ud r^2+\ud x_1{}^2+\ud x_2{}^2+\ud x_3{}^2}{r^2}-\frac{\ud t^2}{r^{2 z_1}}\right]\,,\quad A_1=Q_1 r^{\zeta-z_1}\ud t\,,\quad \phi=\kappa\ln r\,.
\ee
The answer is yes, but the translation-breaking deformations are exponentially suppressed towards the IR, like $e^{-k r}$, see appendix \ref{app:GroundStatesIsThetaZ}. This is in sharp contrast to the previous, anisotropic saddle points, where deformations were all power-like. This is however expected: degrees of freedom at nonzero momentum $\vec k$ should be exponentially suppressed at finite $z_1$ \cite{Hartnoll:2012rj,semilocal}, and this is precisely what we observe. Note that the IR here is $r\to+\infty$,  again differently from the anisotropic case. The behaviour of the translation-breaking modes in \eqref{sol422} and \eqref{solThetaZ1} is extremely reminiscent of the behaviour of Bessel functions for small and large values of their argument, hinting that these two classes of solutions may be complementary to one another.

Note that there are two families of solutions, depending on whether the density sources a marginal ($\zeta=\theta-3$) or irrelevant (power-like) deformation ($z_1=1$).

An obvious second step is to check the existence of semi-locally critical solutions, corresponding to the formal limit $z_1\to+\infty$, $\tilde\theta=\theta/z_1$ finite, of the solutions \eqref{solThetaZ1}. We thus find solutions conformal to AdS$_2\times R^3$
\be\label{solSL1}
\ud s^2=r^{\frac23\tilde\theta}\left(\ud x_1{}^2+\ud x_2{}^2+\ud x_3{}^2+\frac{L^2\ud r^2-\ud t^2}{r^2}\right),\quad A_1=Q_1r^{\tilde\theta-1}\ud t\,,\quad\phi=\kappa\ln r\,.
\ee
There are two families, depending on whether the magnetic field is marginal or irrelevant in the IR:
\be
A_2= r^{\frac\psi2}\left(Q_2+Q_{21}r^{\alpha_6^-}\right)\omega_2\,.
\ee
In the marginal case, $Q_2\neq0$, while in the irrelevant case, $Q_2=0$ and the stress-tensor from $A_2$ is subleading in the field equations.
Both families are detailed in appendices \ref{app:GroundStatesIsAdS2} and \ref{app:GroundStatesIsSL}.

The deformations are all power-like, and can depend on the momentum along the helix director, as already observed for finite momentum deformations of semi-locally critical states \cite{Hartnoll:2012rj,semilocal}. Moreover, there can be a region of dynamical instability, see figure \ref{Fig:k-instability}.

\subsection{Low temperature behaviour of the DC conductivity\label{subsection:HelicalIsDC}}

\paragraph{Isotropic saddle points \protect\eqref{solThetaZ1}\\}
The ground states \eqref{solThetaZ1} display, perhaps unsurprisingly, an exponentially-enhanced DC conductivity at low temperatures, swamping out completely the quantum critical term (which goes like a power of $T$). As a consequence, we expect them to describe metals with a sharp Drude peak.
\be
\sigma_{DC}\sim \frac{T^{\frac{z_1-1}{z_1}}}{\bar k^2}e^{2\bar k T^{-1/z_1}}+\cdots,\qquad \bar k=L k
\ee
which is consistent with similar results for spectral weights at finite $k$ and $z_1$ obtained in \cite{semilocal}.

\paragraph{Semi-locally critical saddle points \protect\eqref{solSL1}\\}
On the other hand, taking an infinite $z_1$ limit, the semi-locally critical ground states \eqref{solSL1} display power-like DC conductivities. For the solution with a marginally relevant magnetic field,
\be
\sigma_{DC}\sim \frac{T^{\tilde\theta}}{k^2}+\frac{Q^2}{Q_2{}^2k^2} T^{\tilde\theta}+\cdots
\ee
The quantum critical term and the magnetic term have the same scaling with temperature at leading order. Positivity of the specific heat implies $\tilde\theta<0$, so the system is always a metal. If the second term is enhanced (small breaking of translation symmetry), we can expect a sharp Drude peak. Note that as seen in other semi-locally critical setups, a linear resistivity implies a linear heat-capacity/entropy \cite{DSZ}.

Turning to the saddle point \eqref{solSL1}  with an irrelevant magnetic field, we find that the dominant contribution to the DC conductivity comes from the magnetic, dissipative piece
\be
\sigma_{DC}\sim \frac{Q^2}{Q_2{}^2k^2}T^{\theta+\alpha_6^-}\left(1+O(T^{-\alpha_6^-})\right)+\cdots
\ee
Here, the IR is $r\to+\infty$, so $\alpha_6^-$ is negative when irrelevant, and then indeed the subleading pieces decay as the temperature decreases. By the same argument as above and given positivity of the specific heat, the system is always metallic and expected to display a sharp Drude peak.

\acknowledgments
\label{ACKNOWL}

We would like to thank M.~Blake, R.~Davison, S.~Hartnoll, J.~P.~Gauntlett, J.~Ren, D.~van der Marel, D.~Vegh, B.~Withers, J.~Zaanen for stimulating discussions.

This work was supported in part by European Union's Seventh Framework Programme under grant agreements (FP7-REGPOT-2012-2013-1) no 316165,
PIF-GA-2011-300984, the EU program ``Thales'' MIS 375734, by the European Commission under the ERC Advanced Grant BSMOXFORD 228169 and was also co-financed by the European Union (European Social Fund, ESF) and Greek national funds through the Operational Program ``Education and Lifelong Learning'' of the National Strategic Reference Framework (NSRF) under ``Funding of proposals that have received a positive evaluation in the 3rd and 4th Call of ERC Grant Schemes''. This work was also partially supported by the People Programme (Marie Curie Actions) of the European Union's FP7 Programme under REA Grant Agreement No 317089.  We also thank the ESF network Holograv for partial support.

\newpage
\appendix
\renewcommand{\theequation}{\thesection.\arabic{equation}}
\addcontentsline{toc}{section}{Appendix}
\section*{Appendices}

\section{The equations of motion \label{app:eoms}}

In this appendix we provide the detailed equations of motion in the helical Ansatz.
The action is
 \be\label{1Action}
 S=M^{3}\int d^{5}x\sqrt{-g}\left[R-{1\over 2}(\partial\phi)^2+V(\phi)-\frac{Z_1(\phi)}4F_1^2-\frac{Z_2(\phi)}4F_2^2\right]
\ee
The Ansatz for the metric is
\be\label{1Metric}
	\ud s^2=-D(r)\ud t^2+B(r)\ud r^2+C_1(r)\omega_1^2+C_2(r)\omega_2^2+C_3(r)\omega_3^2
\ee
where
\be\label{1Bianchi}
	\omega_1=\ud x_1\,,\quad \omega_2=\cos(kx_1)\ud x_2+\sin(kx_1)\ud x_3\,,\quad \omega_3=\sin(kx_1)\ud x_2-\cos(kx_1)\ud x_3
\ee
are the three Bianchi VII$_0$ one-forms. The Ansatz for the scalar and the gauge fields read:
\be\label{1Gauge}
	\phi=\phi(r)\,,\quad A_1=A_1(r)\ud t\,,\quad A_2=A_2(r)\omega_2
\ee
The Maxwell equation for $A_1$ reads
\be\label{1M1}
\sqrt{\frac{1}{BC_1C_2C_3D}}\left(Z_1(\phi)\sqrt{\frac{C_1C_2C_3}{BD}}A_1'\right)'=0
\ee
The Maxwell equation for $A_2$ reads
\be\label{1M2}
 \sqrt{\frac1{BC_1C_2C_3D}}\left(Z_2(\phi)\sqrt{\frac{C_1C_3D}{C_2B}}A_2'\right)'=\frac{k^2A_2}{C_1C_3}Z_2(\phi)
\ee
The scalar equation reads:
\be\label{1ScalEq}
 0=\sqrt{\frac{B}{C_1C_2C_3D}}\left(\frac{\phi'\sqrt{C_1C_2C_3D}}{\sqrt{B}}\right)'+BV'(\phi)+\frac{Z_1'(\phi)(A_1')^2}{2D}-\frac{Z_2'(\phi)(A_2')^2}{2C_2}-\frac{Z_2'(\phi)(kA_2)^2B}{2C_1C_3D}
\ee

The Einstein equations can be written as follows. Denote by ${\cal E}_{\m\n}$ the equation $R_{\m\n}-{1\over 2}g_{\m\n}R-T_{\m\n}=0$. Then,
${\cal E}_{rr}$ is
\be
-{1\over 2}\phi'^2-BV+\left({Z_2A_2^2\over C_3}+{C_2\over C_3}+{C_3\over c_2}-2\right){k^2B\over 2C_1}+{Z_1A_1'^2\over 2D}-{Z_2A_2'^2\over 2C_2}+
\label{e1}\ee
$$+
{D'\over 2D}\left({C_1'\over C_1}+{C_2'\over C_2}+{C_3'\over C_3}\right)+{1\over 2}\left({C_1'\over C_1}{C_2'\over C_2}+{C_1'\over C_1}{C_3'\over C_3}+{C_3'\over C_3}{C_2'\over C_2}\right)=0
$$
where primes denote derivatives with respect to $r$.

The combination $-{\cal E}_{rr}-{\cal E}_{tt}$ gives
\be
{C_1''\over C_1}+{C_2''\over C_2}+{C_3''\over C_3}+\phi'^2-{1\over 2}\left({D'\over D}+{B'\over B}\right)\left({C_1'\over C_1}+{C_2'\over C_2}+{C_3'\over C_3}\right)-{1\over 2}\left({C_1'^2\over C_1^2}+{C_2'^2\over C_2^2}+{C_3'^2\over C_3^2}\right)+{Z_2A_2'^2\over C_2}=0
\label{e2}\ee

The combination ${\cal E}_{tt}+{\cal E}_{11}$ gives
\be
\left(2-{Z_2A_2^2\over C_3}-{C_2\over C_3}-{C_3\over C_2}\right){k^2B\over C_1}-{Z_1A_1'^2\over D}+{C_1'\over 2C_1}\left({C_1'\over C_1}-{C_2'\over C_2}-{C_3'\over C_3}+{B'\over B}\right)+
\label{e3}\ee
$$+{D'\over 2D}\left(-{D'\over D}+{C_2'\over C_2}+{C_3'\over C_3}-{B'\over B}\right)
-{C_1''\over C_1}+{D''\over D}=0
$$

The combination
${\cal E}_{tt}+{1\over 2C_2}\left[{\cos(2kx_1)+1\over \cos(2kx_1)}{\cal E}_{22}+{\cos(2kx_1)-1)\over \cos(2kx_1)}{\cal E}_{33}\right]$ gives
\be
\left({C_2\over C_3}-{C_3\over C_2}\right){k^2B\over C_1}-{Z_1A_1'^2\over D}-{Z_2A_2'^2\over C_2}
+{C_2'\over 2C_2}\left(-{C_1'\over C_1}+{C_2'\over C_2}-{C_3'\over C_3}+{B'\over B}\right)+
\label{e4}\ee
$$+{D'\over 2D}\left(-{D'\over D}+{C_1'\over C_1}+{C_3'\over C_3}-{B'\over B}\right)
-{C_2''\over C_2}+{D''\over D}=0
$$
The combination
${\cal E}_{tt}+{1\over 2C_3}\left[{\cos(2kx_1)-1\over \cos(2kx_1)}{\cal E}_{22}+{\cos(2kx_1)+1)\over \cos(2kx_1)}{\cal E}_{33}\right]$ gives
\be
\left(-{C_2\over C_3}+{C_3\over C_2}-{Z_2A_2^2\over C_3}\right){k^2B\over C_1}-{Z_1A_1'^2\over D}
+{C_3'\over 2C_3}\left(-{C_1'\over C_1}-{C_2'\over C_2}+{C_3'\over C_3}+{B'\over B}\right)+
\label{e5}\ee
$$+{D'\over 2D}\left(-{D'\over D}+{C_1'\over C_1}+{C_2'\over C_2}-{B'\over B}\right)
-{C_3''\over C_3}+{D''\over D}=0
$$
Finally a linear combination using also ${\cal E}_{23}$ gives
\be
{C_3''\over C_3}-{C_2''\over C_2}+{1\over 2}\left(-{B'\over B}+{D'\over D}+{C_1'\over C_1}+{C_2'\over C_2}+{C_3'\over C_3}\right)
\left(-{C_2'\over C_2}+{C_3'\over C_3}\right)-{Z_2a_2'^2\over C_2}+
\label{e6}\ee
$$
+\left(2{C_2\over c_3}-2{C_3\over C_2}+{Z_2a_2'^2\over c_3}\right){k^2B\over C_1}=0
$$
Equation (\ref{e6}) however can be obtained by subtracting (\ref{e5}) from (\ref{e4}).
We will ignore therefore (\ref{e6}) from now on.

The Einstein equations above can also be written in a condensed form as
\bea 0&=&\phi'^2+Z_2(\phi)\frac{(A'_2)^{2}}{C_2}+\sqrt{\frac{BD}{C_1}}\left(\frac{C_1'}{\sqrt{BC_1D}}\right)'+
\sqrt{\frac{BD}{C_2}}\left(\frac{C_2'}{\sqrt{BC_2D}}\right)'+\sqrt{\frac{BD}{C_3}}\left(\frac{C_3'}{\sqrt{BC_3D}}\right)'\label{app:E1}\\
	 0&=&BV(\phi)-\sqrt{\frac{B}{C_1C_2C_3D}}\left(\sqrt{\frac{C_1}{B}}\left(\sqrt{C_2C_3D}\right)'\right)'\label{app:E2}\\
	 0&=&\frac{2Bk^2}{C_1C_2}(C_3-C_2)-\sqrt{\frac{B}{DC_2C_3}}\left(\sqrt{\frac{DC_2}{BC_3}}C_3'\right)'+
\sqrt{\frac{B}{DC_2C_1}}\left(\sqrt{\frac{DC_2}{BC_1}}C_1'\right)'\label{app:E3}\\
	 0&=&(\phi')^2+\frac{k^2A_2^2B}{C_1C_3}Z_2(\phi)-\frac{D'}{D}\left(\frac{C_1'}{C_1}-\frac{C_2'}{C_2}-
\frac{C_3'}{C_3}\right)+\sqrt{\frac{B}{C_1C_2C_3^{1/3}D}}\left(\frac{C_2}{C_3}+2\right)\left(\frac{C_1'
\sqrt{C_3^{1/3}C_2D}}{\sqrt{BC_1}}\right)'\nn\\
	 &&-\sqrt{\frac{B}{C_1^{1/3}C_2C_3D}}\left(\frac{C_2}{C_3}-1\right)\left(\frac{C_3'\sqrt{C_1^{1/3}C_2D}}
{\sqrt{BC_3}}\right)'-\frac{C_1'C_2'}{C_1C_2}-\frac{C_3'C_2'}{C_3C_2}\label{app:E4}\\
	 0&=&\frac{(A_1')^2}{D}Z_1(\phi)-(\phi')^2+\frac{C_1'C_2'}{C_1C_2}+\frac{C_3'C_2}{C_3C_2}-
\frac{C_1'C_3'}{C_1C_3}+\frac12\sqrt{\frac{BC_1}{C_2C_3D}}\left(\frac{C_2}{C_3}-1\right)
\left(\frac{C_3'\sqrt{C_2D}}{\sqrt{BC_1C_3}}\right)'\nn\\
	 &&-\frac12\sqrt{\frac{BC_3}{C_1C_2D}}\left(\frac{C_2}{C_3}+3\right)\left(\frac{C_1'\sqrt{C_2D}}{\sqrt{BC_1C_3}}\right)' -\sqrt{\frac{BC_1C_2C_3}{D}}\left(\frac{D'}{\sqrt{BC_1C_2C_3D}}\right)'\label{app:E5}
\eea

To extract generic scaling ground states, we  make a  generic scaling Ansatz that respects the Bianchi symmetries as in (\ref{1Metric}), (\ref{1Gauge}),
\be\label{ScalingAnsatz}
	\begin{split}
	&A_1(r)=Q_1r^{a_0}+\cdots,\quad A_2(r)=Q_2r^{a_2}+\cdots,\quad \phi(r)=\kappa\log r+\cdots\,, B(r)=B_0 r^{b_0}+\cdots,\\
	& C_1(r)=r^{c_1}+\cdots,\quad C_2(r)=r^{c_2}+\cdots,\quad C_3(r)=\lambda r^{c_3}+\cdots,\quad D(r)=r^{d_0}+\cdots.
	\end{split}
\ee

Introducing it in the field equations, we obtain equations which have to be made algebraic. Some of them yield straightforward constraints which cannot be evaded at leading order. Thus, equations \eqref{app:E1}, \eqref{app:E2}, \eqref{app:E3}, \eqref{app:E4}, \eqref{app:E5}, \eqref{1M1} and \eqref{1M2} give respectively:
\be\label{E12}
\begin{split}
0=& 2 \kappa ^2+c_1^2+c_2^2-2 c_3-b_0 c_3+c_3^2-c_3 d_0-c_1 \left(2+b_0+d_0\right)\\
&-c_2 \left(2+b_0+d_0\right)+2r^{2 a_2-c_2+\kappa  \gamma _2} a_2^2 Q_2^2
\end{split}
\ee
\be\label{E22}
0=-\left(c_2+c_3+d_0\right) \left(-2-b_0+c_1+c_2+c_3+d_0\right)+4 r^{2-\delta  \kappa +b_0} B_0 V_0
\ee
\be \label{E32}
0=4 k^2 \lambda  B_0-4 k^2 r^{c_2-c_3} B_0+r^{-2-b_0+c_1+c_2-c_3} \left(c_1-c_3\right) \left(c_1+c_2+c_3+d_0-2-b_0\right)
\ee
\be\label{E42}
\begin{split}
0=&r^{c_2-c_3} \left(c_3-c_1\right) \left(-2-b_0+c_1+c_2+c_3+d_0\right)-2 k^2 r^{2+2 a_2+b_0-c_1-c_3+\kappa  \gamma _2} B_0 Q_2^2\\
&+\lambda  \left(c_3 \left(2+b_0+c_2+d_0\right)-2 \kappa ^2-2 c_1^2+c_1 \left(4+2 b_0-c_3\right)-c_3^2+2 c_2 d_0\right)
\end{split}
\ee
\be\label{E52}
\begin{split}
0=&\frac1\lambda r^{c_2-c_3} \left(-c_1+c_3\right) \left(-2-b_0+c_1+c_2+c_3+d_0\right)+4 r^{2 a_0-d_0+\kappa  \gamma _1}   a_0^2 Q_1^2\\
&+  c_1 \left(6+3 b_0+c_2-d_0\right)-4 \kappa ^2-3 c_1^2-c_3^2+4 d_0+2 b_0 d_0+2 c_2 d_0-2 d_0^2+c_3 \left(2+b_0+3 c_2+d_0\right)
\end{split}
\ee
\be\label{M12}
0=a_0 Q_1 \left(-2+2 a_0-b_0+c_1+c_2+c_3-d_0+2 \kappa  \gamma _1\right)
\ee
\be\label{M22}
0=-\frac{2 k^2 r^{2+b_0-c_1+c_2-c_3} B_0 Q_2}{\lambda }+a_2 Q_2 \left(-2+2 a_2-b_0+c_1-c_2+c_3+d_0+2 \kappa  \gamma _2\right)
\ee

The analysis of solutions of these leading equations provides a classification of the IR fixed point solutions, which we detail in sections \ref{section:HelicalAn} and \ref{section:HelicalIs} and appendix \ref{app:GroundStatesIs}.

\section{Null Energy Condition \label{app:NEC}}

Given the metric Ansatz
\be
\ud s^2=-D(r)\ud t^2+B(r)\ud r^2+C_1(r)\omega_1^2+C_2(r)\omega_2^2+C_3(r)\omega_3^2\,,
\ee
we can identify a null vector
\be
\begin{split}
N^\mu=&\left(\frac1{\sqrt{D(r)}},\frac{c_r}{\sqrt{B(r)}},\frac{c_1}{\sqrt{C_1(r)}},\frac{c_2}{\sqrt{C_2(r)\cos(kx)^2+C_3(r)\sin(kx)^2}},\right.\\
&\qquad\qquad\qquad\qquad\qquad\qquad\qquad\left.\frac{c_3}{\sqrt{C_3(r)\cos(kx)^2+C_2(r)\sin(kx)^2}}\right).
\end{split}
\ee
The Null Energy Condition is a condition such that the geometry is supported by 'reasonable' matter content and reads
\be
T_{\mu\nu}N^\mu N^\nu\geq0
\ee
which via Einstein's equations translates into a conidtion on the geometry
\be
G_{\mu\nu}N^\mu N^\nu\geq0
\ee
where $G_{\mu\nu}$ is the Einstein tensor. As there are four independent coefficients $(c_r,c_1,c_2,c_3)$, this yields four inequalities on the metric functions
\be\label{NEC1}
\begin{split}
0\leq&\sum_{i=1}^3\frac{B'C_i'}{B^2C_i}+\frac{D'C_i'}{BC_iD}-\frac{2C_i''}{BC_i}+\frac{C_i'^2}{BC_i^2}\\
0\leq&-\frac{2k^2C_2}{C_1C_3}-\frac{2k^2C_3}{C_1C_2}+\frac{4k^2}{C_1}-\frac{2C_1''}{BC_1}+\frac{2D''}{BD}+\frac{C_1'^2}{BC_1^2}-\frac{D'^2}{BD^2} +\frac{B'C_1'}{B^2C_1}+\frac{D'C_3'}{BC_3D}\\
&\qquad+\frac{D'C_2'}{BC_2D}-\frac{C_1'C_2'}{BC_1C_2}-\frac{C_1'C_3'}{BC_1C_3}-\frac{B'D'}{B^2D}\\
0\leq&\frac{2k^2C_3^2}{C_1C_2}-\frac{2k^2C_2}{C_1}-\frac{2C_3''}{B}+\frac{2C_3D''}{BD}+\frac{C_3D'C_1'}{BC_1D}+\frac{C_3D'C_2'}{C_2DB}-\frac{C_2'C_3'}{BC_2}-\frac{C_3B'D'}{B^2D}\\
&\qquad -\frac{C_3D'^2}{BD^2}+\frac{C_3'^2}{BC_3}-\frac{C_1'C_3'}{BC_1}+\frac{B'C_3'}{B^2}\\
0\leq&\frac{2k^2C_2^2}{C_1C_3}-\frac{2k^2C_3}{C_1}-\frac{2C_2''}{B}+\frac{2C_2D''}{BD}+\frac{C_2D'C_1'}{BC_1D}+\frac{C_2D'C_3'}{C_3DB}-\frac{C_3'C_2'}{BC_3}-\frac{C_2B'D'}{B^2D}\\
&\qquad -\frac{C_2D'^2}{BD^2}+\frac{C_2'^2}{BC_2}-\frac{C_1'C_2'}{BC_1}+\frac{B'C_2'}{B^2}
\end{split}
\ee

Evaluated on the geometries \eqref{sol422}, at leading order in $r$, these inequalities translate into:
\be\label{NEC2}
\begin{split}
0\leq&-2-2z_2{}^2-2z_2+\left.\theta ^2\right/3+2z_1+2z_1z_2-z_1\theta\\
0\leq&\frac{-\lambda }{2}k^2L^2-\left(z_1-1\right)\left(\theta -2-z_1-2z_2\right) \\
0\leq&\frac{\lambda }{2}k^2L^2-\left(z_1-z_2-1\right)\left(-z_1-2-2z_2+\theta \right) \\
0\leq&\frac{-\lambda }{2}k^2L^2-\left(z_1-z_2\right)\left(\theta -2-z_1-2z_2\right)
\end{split}
\ee
and one may check that they are not violated in the allowed parameter space.


\section{Isotropic IR solutions\label{app:GroundStatesIs}}

\subsection{Hyperscaling violating solutions with helical deformations\label{app:GroundStatesIsThetaZ}}

In this section, we describe ground states which are translation invariant with both hyperscaling violation $\theta\neq0$ and anisotropicity between time and space $z_1\neq1$:
\be\label{ThetaZsolns}
\begin{split}
&\ud s^2=r^{\frac23(\theta-3)}\left(\omega_1^2+ \omega^2_2+\omega_3^2+L^2\ud r^2\right)-r^{\frac{2}{3} (\theta-3 z_1)}\ud t^2,\\
&A_1=Q_1r^{\zeta-z_1}\ud t\,,\quad  L^2=\frac{(3+z_1-\theta ) (2+z_1-\theta )}{V_0}\,,\quad e^{\phi}=r^{\sqrt{\frac{2}{3}} \sqrt{(\theta-3 ) (3-3 z_1+\theta )}}\\
&\kappa\delta = \frac{2}{3} \theta ,\quad \kappa\gamma _1=3-\zeta-\frac13\theta\,,\quad \kappa =\sqrt{\frac{2}{3}} \sqrt{(\theta-3 ) (3-3 z+\theta )}
\end{split}
\ee
There are two families, depending on whether the density generates a marginal deformation:
\be
\zeta=\theta-3\,,\quad Q_1=\sqrt{\frac{2 (z_1-1)}{3+z_1-\theta }}
\ee
or an irrelevant one
\be
z_1=1\,.
\ee

Now we turn to  modes corresponding to the magnetic field, as well as breaking translation invariance.
Parameterizing the perturbations such that $\delta g_{\omega_2\omega_2}=-\delta g_{\omega_3\omega_3}$, we find exponential deformations
\be\label{ExpDefznot1}
\begin{split}
&\delta g_{\omega_3\omega_3}=e^{-2 k r} r^{\frac{1}{6} (3 z_1+\theta )}\sum_{i\leq 1}c_3^i r^{-i}\\
&\delta A_2=e^{-k r} r^{\frac{1}{6} (3 z_1-\theta -3\kappa \gamma_2)}\sum_{i\leq 0}a_2^i r^{-i}\\
\end{split}
\ee
which are valid when the IR is $r\to+\infty$. It can be checked that these two perturbations backreact at quadratic order on other perturbations with factors of $e^{-4 k r}$, $e^{-2 k r}$ respectively.

The parameter space then reads for a marginal density deformation
\be\label{ParSpaceIsoHVMarginal}
V_0>0\,,\qquad \left(1<z_1\leq 2\,,\quad\theta <-3+3 z_1\right)\quad \|\quad \left(z_1>2\,,\quad\theta <3\right)
\ee
and
\be\label{ParSpaceIsoHVIrr}
V_0>0\,,\qquad \zeta>4\,\quad \|\quad \zeta<\theta-3
\ee
for an irrelevant density deformation.
There are other modes, which can be developed as power series. One is the universal temperature mode $\beta_u=3+z_1-\theta$ dual to a marginal mode. Finally, for a marginal density deformation, there is a pair
\be
\begin{split}
&\beta_\pm=\frac12(3+z_1-\theta)\pm\frac{\sqrt{X}}{2(3z_1-3-\theta)}\\
&X=(3+z_1-\theta ) (3z_1-3-\theta) \left(-57+30 z_1+27 z_1^2+24 \theta -28 z_1 \theta +\theta ^2\right)
\end{split}
\ee
where $\beta_-$ is always irrelevant and $\beta_+$ always relevant for the parameter space \eqref{ParSpaceIsoHVMarginal}. On the other hand, when the density deformation is irrelevant, the electric potential generates a mode that behaves like $O(Q_1^2r^{3+\zeta-\theta})$ at quadratic order in the other fields.


\subsection{AdS$_2\times R^3$ solutions\label{app:GroundStatesIsAdS2}}

An electric AdS$_2\times R^3$ can be found, with the magnetic field and helical deformations turned off in the background solution:
\be
\begin{split}
&B(r)=\frac{e^{\gamma_1\phi_0}}{V_0 r^2}\,,\qquad D(r)=\frac1{r^2}\,,\qquad C_1(r)=C_{10}\,,\qquad C_2(r)=C_3(r)=\lambda\,,\\
&A_1(r)=\sqrt{2} e^{-\frac{1}{2} \gamma _1 \phi _0}\frac1r\,,\qquad A_2(r)=0\,,\qquad \phi(r)=\phi_0\,,\qquad \delta=\gamma_1\,.
\end{split}
\ee
Note that we have assumed the form \eqref{IRcouplings} for the scalar potential and gauge couplings, and that the solution only exists for the relation $\delta=\gamma_1$. This can be alleviated by making no assumption such as \eqref{IRcouplings} but rather keeping the couplings generic and assuming the scalar field extremizes the effetive scalar potential at some finite value $\phi_\star$, along the lines developed in \cite{gk2012}. The expressions become a little cumbersome but qualitatively similar. For instance, the relation above becomes $V'_\star Z_{1,\star} = Z'_{1,\star}V_\star$ where the couplings are evaluated at $\phi_\star$.

We now consider  the deformations. We expect 12 modes, coming in pairs that sum to $1$. For brevity we only give the expressions for the irrelevant modes $i=1\dots6$, their conjugates can be worked out from their sum.
\be
\begin{split}
&B(r)=\frac{e^{\gamma_1\phi_0}}{V_0 r^2}\left[1+\sum_i c_i^B r^{\alpha_i}\right],\qquad \phi(r)=\phi_0+\sum_i c_i^2 r^{\alpha_i},\qquad i=1\dots5\\
&D(r)=\frac{1}{r^2}\left[1+\sum_i c_i^D r^{\alpha_i}\right],\quad C_1(r)=C_{10}\left[1+\sum_i c_i^1 r^{\alpha_i}\right],\\
&C_2(r)=\lambda\left[1+\sum_i c_i^2 r^{\alpha_i}\right],\quad C_3(r)=\lambda\left[1+\sum_i c_i^3 r^{\alpha_i}\right],\\
&A_1(r)=\sqrt{2} e^{-\frac{1}{2} \gamma _1 \phi _0}\frac1r\left[1+\sum_i c_i^e r^{\alpha_i}\right]\,,\qquad A_2(r)= c_6^a r^{\alpha_6}\,.
\end{split}
\ee
The magnetic deformations $c^a_{6}$ and $\alpha_{6}$ decouple from the others at linear order.

First, there are three marginal modes $\alpha_{1,2,3}=0$ with $c_1^D$, $c^1_2$ and $c^3_3$ turned on which corresponds to rescalings of time, $C_{10}$ and $\lambda$.

Then there are three irrelevant modes. One is associated with the magnetic field with $c_6^m$ turned on and
\be
\alpha_{6}=\frac{1}{2}-\sqrt{\frac14+L^2k^2}
\ee
as well as two others
\be
\alpha_4=-1\,,\qquad \alpha_5=\frac{1}{2}-\sqrt{\frac14+\frac{4 k^2 L^2}{C_{10}}}
\ee
with $c_4^B$ and $c^2_5$ turned on respectively. Note that $\alpha_{4,5,6}$ are always irrelevant, contrary to the results of \cite{Donos:2012js}, where it was essential to turn on a Chern-Simons term to trigger a metal/insulator transition. In the next appendix \ref{app:GroundStatesIsSL}, we shall see that allowing for a running scalar can allow for instabilities, where a mode becomes relevant when varying the helix periodicity $k$.


\subsection{Semi-locally critical solutions\label{app:GroundStatesIsSL}}

In this section, we write down solutions where the metric is conformal to AdS$_2\times R^3$, which means they are semi-locally critical, \cite{cgkkm,semilocal}. We find two families, which are distinguished by whether the magnetic field appears as a marginal or an irrelevant deformation. Then we also work out their static, purely radial deformations. In any case, the solutions are written
\be\label{SemiLocal}
\begin{split}
&B(r)=L^2 r^{\frac{2\tilde\theta}{3}-2}\left[1+\sum_i c_i^B r^{\alpha_i}\right],\quad L^2=\frac{(1-\tilde\theta)^2}{V_0}\,,\quad D(r)=r^{\frac{2\tilde\theta}{3}-2}\left[1+\sum_i c_i^D r^{\alpha_i}\right]\\
&C_j(r)= r^{\frac{2\tilde\theta}{3}}\left[1+\sum_i c_i^j r^{\alpha_i}\right],\qquad j=1,2,3\,,\\
&\phi(r)=\kappa\left[\log r +\sum_i c_i^\phi r^{\alpha_i}\right],\quad A_1(r)=Q_1r^{\tilde\theta-1}\left[1+\sum_i c_i^{e} r^{\alpha_i}\right]\\
&A_2(r)= r^{\frac\psi2}\left(Q_2+\sum_i c_i^{m} r^{\alpha_i}\right),\qquad \gamma_1=-2\delta\,.
\end{split}
\ee
These solutions are of interest, since time and space scaling under rigid scale transformations are decoupled. Consequently, degrees of freedom can be created at finite momentum for no cost \cite{semilocal}. Formally, they correspond to taking a limit $\theta\to+\infty$, $z_1\to+\infty$ in the solutions \eqref{ThetaZsolns} while keeping their ratio finite, where $z_1$ parametrizes how time scales. Note that they exist only if $\gamma_1=-2\delta$.

\paragraph{Marginal magnetic field\\}
The rest of the solution reads
\be\label{SemiLocalMagRel}
\begin{split}
	&Q_1^2=\frac{4 \tilde\theta ^2-6 \delta ^2 \left(-3+\tilde\theta ^2\right)}{9 \delta ^2 (-1+\tilde\theta )^2}\,,\qquad Q_2^2=\frac{4 \left(6 \delta ^2 (-3+\tilde\theta )-4 \tilde\theta \right) \tilde\theta }{9 \delta ^2 (-1+\tilde\theta )^2}\,,\\
&\gamma _2= \frac{3 \delta -\delta  \tilde\theta }{2 \tilde\theta }\,,\qquad \kappa = \frac{2 \tilde\theta }{3 \delta }\,,\qquad \psi=\tilde\theta-1
\end{split}
\ee
$\g_2$ and $\delta$ are not fixed a priori but should be in a range where both $Q_{1,2}$ are real:
\be \label{Constraints}
\begin{split}
&\gamma _2<-\frac{1}{\sqrt{6}}\,,\qquad-\frac{1}{3 \gamma _2}<\delta <\gamma _2+\sqrt{1+3 \gamma _2^2}\qquad \|\\
&\gamma _2>\frac{1}{\sqrt{6}}\,,\qquad \gamma _2-\sqrt{1+3 \gamma _2^2}<\delta <-\frac{1}{3 \gamma _2}
\end{split}
\ee
Given the range \eqref{Constraints}, the IR can be determined to be $r\to+\infty$, as this is where the metric ($t,x_1,x_2,x_3$) elements vanish. Moreover the gauge fields also vanish there.

The $\alpha_i$ characterizing the deformations come in pairs summing to $1-\tilde\theta$ with $i=1\dots12$. Below, we only give the potentially irrelevant deformations $i=1\dots6$, their conjugates can then be deduced. There are two marginal modes $\alpha_1=\alpha_2=0$, which are a rescaling of time and a constant shift of the scalar field with $c_1^D$ and $c_2^\phi$ the independent amplitudes respectively. Then, there are four more modes which can be irrelevant, but they are solutions to a polynomial of order 8 and their analytic expression is intricate. Depending on the parameter space some can be complex, which points to a dynamical instability, though the modes do not depend on the value of the helix periodicity $k$. We give below the expression of the polynomial for the modes
\be
\begin{split}
&0=\sum_{i=0}^8 X_i\beta^i\,,\quad X_8=243 \delta ^4\,,\quad X_7=972 \delta ^4 (-1+\tilde\theta )\,,\\
&X_6=9 \left(-8 \delta ^2 \tilde\theta ^2+9 \delta ^6 \left(-9-6 \tilde\theta +7 \tilde\theta ^2\right)+3 \delta ^4 \left(27-78 \tilde\theta +35 \tilde\theta ^2\right)\right)\,,\\
&X_5=27 \delta ^2 \left(8 (1-\tilde\theta ) \tilde\theta ^2+\delta ^2 \left(45-63 \tilde\theta +39 \tilde\theta ^2-21 \tilde\theta ^3\right)+9 \delta ^4 \left(9-3 \tilde\theta -13 \tilde\theta ^2+7 \tilde\theta ^3\right)\right)\\
&X_4=-16 \tilde\theta ^4-36 \delta ^2 \tilde\theta ^2 \left(7-6 \tilde\theta +3 \tilde\theta ^2\right)-54 \delta ^4 \left(27-66 \tilde\theta +94 \tilde\theta ^2-66 \tilde\theta ^3+19 \tilde\theta ^4\right)\\
&\qquad\qquad+27 \delta ^6 \left(-135+180 \tilde\theta +6 \tilde\theta ^2-108 \tilde\theta ^3+41 \tilde\theta ^4\right)\\
&X_3=-32 (-1+\tilde\theta ) \tilde\theta ^4+144 \delta ^2 \tilde\theta ^2 \left(1+\tilde\theta -3 \tilde\theta ^2+\tilde\theta ^3\right)\\
&\qquad\qquad+27 \delta ^4 \left(-9+93 \tilde\theta -130 \tilde\theta ^2+50 \tilde\theta ^3-5 \tilde\theta ^4+\tilde\theta ^5\right)\\
&\qquad\qquad-27 \delta ^6 \left(-135+315 \tilde\theta -318 \tilde\theta ^2+222 \tilde\theta ^3-107 \tilde\theta ^4+23 \tilde\theta ^5\right)\\
&X_2=-6 (-1+\tilde\theta )^2 \left(-8 \tilde\theta ^4+6 \delta ^2 \tilde\theta ^2 \left(-7-10 \tilde\theta +5 \tilde\theta ^2\right)\right.\\
&\quad\quad+27 \delta ^6 \left(-27+12 \tilde\theta +22 \tilde\theta ^2-20 \tilde\theta ^3+5 \tilde\theta ^4\right)\\
&\qquad\qquad\left.-9 \delta ^4 \left(9+6 \tilde\theta +46 \tilde\theta ^2-50 \tilde\theta ^3+13 \tilde\theta ^4\right)\right)\\
&X_1=-8 \left(3 \delta ^2 (-3+\tilde\theta )-2 \tilde\theta \right)^2 (-1+\tilde\theta )^3 \left(-2 \tilde\theta ^2+3 \delta ^2 \left(-3+\tilde\theta ^2\right)\right)\\
&X_0=2 \left(3 \delta ^2 (-3+\tilde\theta )-2 \tilde\theta \right)^2 (-1+\tilde\theta )^4 \left(-2 \tilde\theta ^2+3 \delta ^2 \left(-3+\tilde\theta ^2\right)\right)
\end{split}
\ee


\paragraph{Irrelevant magnetic field\\}
In this case, the rest of the solution reads
\be\label{SemiLocalMagIrr}
\begin{split}
&\kappa=\sqrt{\frac23\tilde\theta(\tilde\theta-3)}\,,\qquad Q_1=\sqrt{\frac2{1-\tilde\theta}}\\
&Q_2=0\,, \qquad \delta=\sqrt{\frac{2\tilde\theta}{3(\tilde\theta-3)}}\,,\quad \gamma_2=\frac{2 \tilde\theta -3 \psi }{\sqrt{6(\tilde\theta-3)\tilde\theta}}\,.
\end{split}
\ee

 The background solution implies that $\tilde\theta(\tilde\theta-3)>0$ in order to be well-defined. Since the entropy density reads $S\sim T^{-\tilde\theta}$ as usual, the local stability condition implies $\tilde\theta<0$, which means that the IR is $r\to+\infty$.

Only the marginal or irrelevant deformations are given. $\alpha^-_{1\dots6}$ have conjugates $\alpha^+_{1\dots6}$ which sum to $1-\tilde\theta$. There are three marginal deformations $\alpha^-_{1,2,3}=0$ which match to rescalings of time, $C_{10}$ and $\lambda$ with $c_1^D$, $c_2^1$ and $c_3^2=c_3^3$ turned on. Their conjugates $\alpha^+_{1,2,3}=1-\tilde\theta$ are always relevant in the IR, as they should be.

Then there are two other deformations with $c_4^B$ and $c^2_5=-c^3_5$ turned on.
\be
\alpha^\pm_4=\frac{1-\tilde\theta}{2}\left(1\pm\sqrt{\frac{\tilde\theta-27}{\tilde\theta-3}}\right),\quad \alpha^\pm_5=\frac{1-\tilde\theta}2\pm\sqrt{\frac{(1-\tilde\theta)^4}{4}+\frac{4k^2L^2}{C_{10}}}
\ee

The last deformation comes from the magnetic field. To determine its dimension, it is necessary to backreact it on the other fields and find the subleading behaviour it generates:
\be
\alpha_6^\pm=\frac12(1-\tilde\theta)\pm\sqrt{\frac{ k^2 L^2}{C_{10}}+\frac14(1-\tilde\theta +\psi )^2}
\ee

$\alpha_4^+$ is always relevant, while $\alpha_4^-$ is irrelevant for $\tilde\theta<0$, relevant otherwise and even complex if $3<\tilde\theta<27$. So let us restrict to $\tilde\theta<0$, for which the IR is $r\to+\infty$. Since $1-\tilde\theta>0$, all the $\alpha_i^+$ are always positive and so relevant.  Then the $\alpha_i^-$ need to be negative to be irrelevant.

For $\tilde\theta<0$, $\alpha_5^+$ is always relevant, $\alpha_5^-$ always irrelevant.

\FIGURE{
\includegraphics{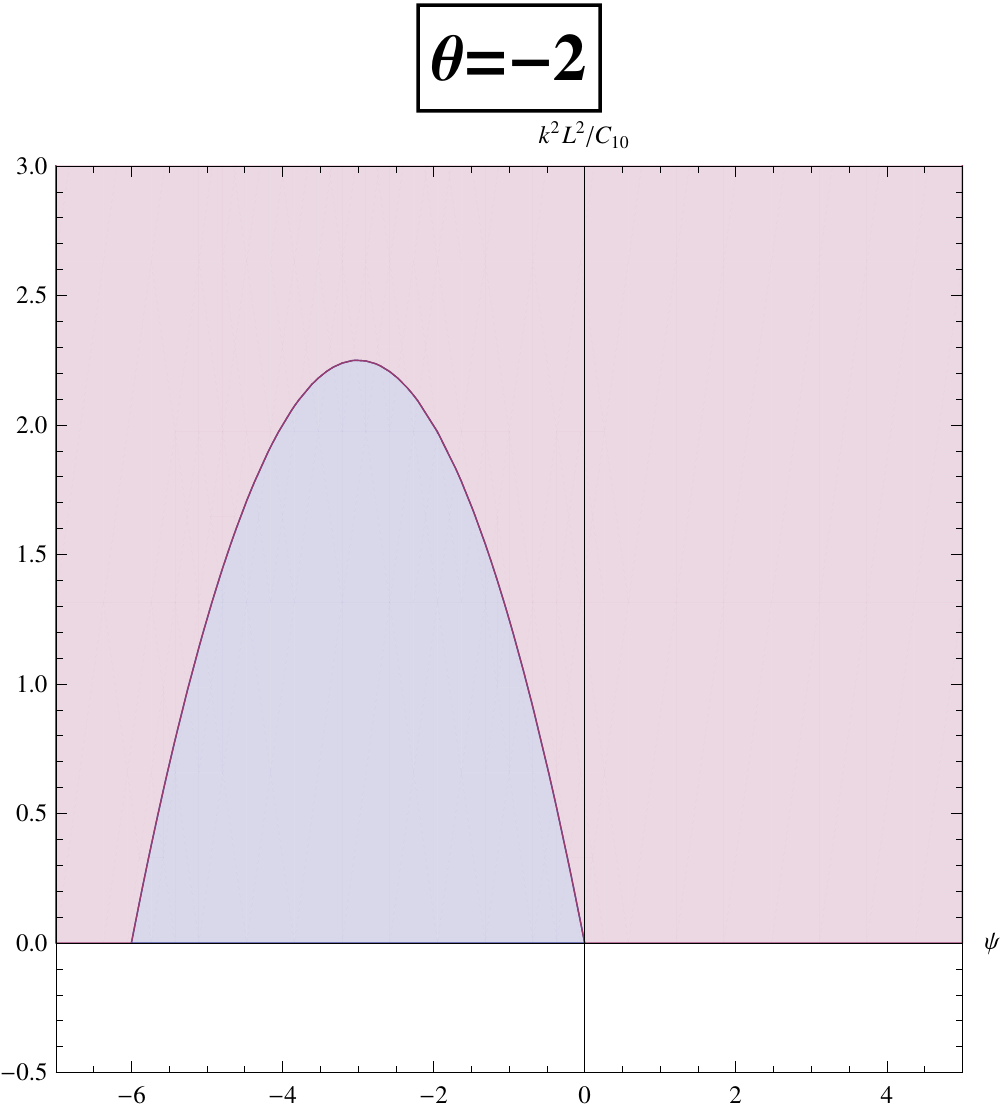}
\caption{RG-stability of the solution \protect\eqref{SemiLocal} with an irrelevant deformation $\alpha_6^-<0$ sourced by the magnetic field. In purple, region of stability ($\alpha_6^-<0$) and in blue, the region of instability ($\alpha_6^->0$) for a representative value of $\tilde\theta<0$. On the horizontal axis is $\psi$, on the vertical axis is the ratio $k^2L^2/C_{10}$. $C_{10}$ can be tuned by UV data in order to destabilize the solution.}
\label{Fig:k-instability}
}

Finally, $\alpha_6^-$ can become relevant at small enough $k^2L^2/C_{10}$ if
\be
0<\frac{ k^2 L^2}{C_{10}}<-\frac{1}{4} \psi  (2-2 \tilde\theta +\psi )\,,\qquad \tilde\theta<0\,,\qquad 2 (\tilde\theta-1 )<\psi <0
\ee
This should be achievable by varying the coupling corresponding to $C_{10}$ in the UV,  so that we would find the insulating solution at large $k/\mu$. The constant scalar case is $\tilde\theta=\psi=0$, so that explains why in this case this not possible without the Chern-Simons term, since the condition reduces to $k^2L^2/C_{10}=0$. Also, note that this is not a dynamical instability, since the mode does not become complex.


\section{Anisotropic partially hyperscaling violating IR solutions\label{app:GroundStatesIsPL}}

In this section, we describe solutions which are only partially hyperscaling violating. By this, we mean that they are written
\be\label{PartLocCrit}
\begin{split}
&\ud s^2=r^{\frac23\tilde\theta}\left[-\frac{\ud t^2}{r^{2z}}+\frac{L^2 dr^2+\omega_2^2+\lambda\omega_3^2}{r^2}+\omega_1^2\right],\quad L^2= \frac{(2+z-\tilde\theta )^2}{V_0}\\
&\phi=\kappa\log r\,,\qquad A_2=Q_2 r^{a_2}\omega_2\,,\qquad A_1=Q_1 r^{\tilde\zeta-z}\ud t\,.
\end{split}
\ee
The $x_1$ direction no longer scales under rigid scale transformations, though the transverse directions $(x_2,x_3)$ still do according to \eqref{ScalingPartiallyHV}. Formally, they correspond to the $\theta\to+\infty$, $z_1\to+\infty$, $z_2\to+\infty$, $\zeta\to+\infty$ limit of the solutions \eqref{sol422}, where the ratios $\tilde\theta=\theta/z_2$, $\tilde\zeta=\zeta/z_2$ and $z=z_1/z_2$ are kept finite. Indeed, taking such a limit, the series in powers of $r^2$ vanishes, leaving an exact solution.

Two families are described, depending on whether the density deformation is marginal or irrelevant. The radial, static deformations are straightforward to work out, but are extremely cumbersome, so we will not report them here.

\paragraph{Marginal density deformation\\}

The rest of the solution reads
\be
\begin{split}
&\kappa = \frac{4}{2 \delta +\gamma _1}=\frac{2 \tilde\theta }{3 \delta }\,,\quad  \gamma _1= -\frac{2 \delta  (-3+\tilde\theta )}{\tilde\theta }\,,\quad \gamma _2= \frac{\delta  \left(-3+\tilde\theta -3 a_2\right)}{\tilde\theta }\\
&\tilde\zeta=\tilde\theta-2\,,\qquad Q_1^2=\frac{-6-6 \lambda -2 \tilde\theta ^2 \lambda +3 \kappa ^2 \lambda +\tilde\theta  (3+9 \lambda )-3 (1+3 \lambda ) z+6 \lambda  z^2}{3 \lambda  \left(2-\tilde\theta +z\right){}^2}\\
&k^2=\frac{V_0}{(2+z-\tilde\theta ) (-1+\lambda )}\,,\quad Q_2^2=\frac{2 (2+z-\tilde\theta ) \left(-1+\lambda ^2\right)}{2+z-\tilde\theta -(-1+\lambda ) \lambda  a_2^2}
\end{split}
\ee
and the additional constraints have to be solved
\be
2+z-\tilde\theta +(2+z-\tilde\theta ) (\lambda-1 ) \lambda  a_2+(\lambda-1 ) \lambda  a_2^2=0
\ee
\be
\begin{split}
0=&6 (\tilde\theta-3 )^2 \lambda +9 (\tilde\theta-3 )^2 (\lambda-1 ) \lambda ^2 a_2\\
&+\frac{3}{2} (\lambda-1 ) \lambda  \left(3+15 \lambda -6 \tilde\theta  \lambda -18 \lambda ^2+12 \tilde\theta  \lambda ^2-2 \tilde\theta ^2 \lambda ^2+18 \lambda ^3-12 \tilde\theta  \lambda ^3+2 \tilde\theta ^2 \lambda ^3\right) a_2^2\\
&-\frac{9}{2} (\lambda-1 )^2 \lambda ^2 (\tilde\theta  \lambda-1-3 \lambda ) a_2^3+\frac{\tilde\theta^2}{\delta^2}\left(2 \lambda +3  (\lambda-1 ) \lambda ^2 a_2+ (\lambda-1 )^2 \lambda ^3 a_2^2\right)
\end{split}
\ee


\paragraph{Irrelevant density deformation\\}

The other quantites become
\be
\begin{split}
&\kappa \delta = \frac{2 \tilde\theta }{3}\,,\quad\kappa \gamma _2= \frac{2}{3} \left(\tilde\theta-3 -3 a_2\right),\quad \kappa ^2= \frac{2}{3} \left(6 (z-1) a_2-6+3 z (2-\tilde\theta )+\tilde\theta ^2\right),\\
&\lambda = \frac{2+z-\tilde\theta }{(-3+2 z) (2+z-\tilde\theta )+4 (-1+z) a_2}\,,\qquad Q_2^2= \frac{4 (1-z)}{a_2}\,,\\
& k^2= -\frac{a_2 \left(2+z-\tilde\theta +a_2\right) V_0}{(2+z-\tilde\theta ) \left((2 z-3) (2+z-\tilde\theta )+4 (z-1) a_2\right)}
\end{split}
\ee
and $a_2$ obeying the equation:
\be
\begin{split}
&(-3+2 z)^2 (2+z-\tilde\theta )^2+2 (2+z-\tilde\theta ) \left(16-20 z+7 z^2-2 \tilde\theta +z \tilde\theta \right) a_2\\
&+2 \left(16-18 z+5 z^2-4 \tilde\theta +3 z \tilde\theta \right) a_2^2-4 (-1+z) a_2^3=0
\end{split}
\ee
The electric potential is be turned on through a deformation, and parameterized as
\be
A_1=Q_1 r^{\tilde\zeta-z}\ud t\,,\qquad \kappa \gamma _1= 2-\tilde\zeta -\frac{\tilde\theta }{3}
\ee
which generates a mode on the metric and other fields like
\be
1+\#Q_1^2r^{2+\tilde\zeta -\tilde\theta}
\ee


\section{Dyonic, translation invariant IR solutions\label{section:ExactTransInv}}

For completeness, we also mention a solution with both the electric and the magnetic fields turned on, which however does not break translation symmetry ($k=0$):
\be
\begin{split}
	&\ud s^2=r^{\frac{2\theta}3}\left(\frac{-\ud t^2}{r^{2z}}+\frac{L^2\ud r^2+\ud x^2+\ud y^2+\ud z^2}{r^2}\right),\quad L^2=\frac{(3+z-\theta)(2+z-\theta)}{V_0}\\
	&0=Q_1^2+\frac{Q_2^2}4+\frac{(z-1)V_0}{2(\theta-z-2)}\,,\quad A_1=\frac{LQ_1}{\theta-3-z}r^{\theta-3-z}\ud t\\
	&A_2=Q_2(x_2\ud x_1-x_1\ud x_2)\,,\quad e^\phi=r^{\frac\theta\d}\\
	&\delta^2=\frac{2\theta^2}{3(3-\theta)(3z-3-\theta)}\,,\quad \g_1=\frac{(9-2\theta)\delta}{\theta}\,,\quad 3\g_2+2\g_1+\d=0
\end{split}
\ee

 \section{Linear fluctuation equations for the AC conductivity \label{app:conductivityFluct}}

 In this appendix we will consider the gauge field  perturbations around the helical Ansatz \eqref{3}, \eqref{5}.

We perturb the fields as follows
 \be
 \delta A_1=e^{-i\omega t}b_1(r) \omega_1\sp \delta A_2=e^{-i\omega t}b_2(r) \omega_3
 \ee
 \be
 \delta(ds^2)=e^{-i\omega t}\left[g_1(r)dt\otimes \omega_1+g_2(r)\omega_2\otimes\omega_3\right]
\ee
The perturbation of the gauge field equations gives
\be
{1\over Z_1}\sqrt{BC_1 \over DC_2C_3}\left(Z_1\sqrt{DC_2C_3\over BC_1}b_1'\right)'+\omega^2{B\over D}b_1+{A_1'\over D}\left(g_1'-{C_1'\over C_1}g_1\right)=0
\label{ea1}\ee
\be
{1\over Z_2}\sqrt{BC_3 \over DC_1C_2}\left(Z_2\sqrt{DC_1C_2\over BC_3}b_2'\right)'+\omega^2{B\over D}b_2-k^2{BC_3\over C_1C_2}b_2
-{A_2'\over 2C_2}\left(g_2'-{C_3'\over C_3}g_2\right)
\label{ea2}\ee
$$
-\left({1\over C_2}+{1\over C_3}\right){k^2A_2Bg_2\over 2C_1}+{ik\omega A_2Bg_1\over C_1D}=0
$$

From the perturbation of the ($rx_1$) Einstein equation we obtain
\be
{k\over C_2C_3}\left[(C_2-C_3)g_2'-(C_2'-C_3')g_2-Z_2A_2A_2'g_2\right]+{2i\omega\over D}\left(g_1'-{C_1'\over C_1}g_1+Z_1A_1'b_1\right)+
\label{ea3}\ee
$$+
2kZ_2\left({A_2b_2'\over C_3}-{A_2'b_2\over C_2}\right)=0
$$

From the perturbation of the ($tx_1$) Einstein equation we obtain
\be
g_1''+\left({C_2'\over C_2}+{C_3'\over C_3}-{C_1'\over C_1}-{D'\over D}-{B'\over B}\right){g_1'\over 2}+
{ik\omega B\over 2}\left({1\over C_3}-{1\over C_2}\right)g_2+Z_1A_1'b_1'+{ik\omega B A_2Z_2\over C_3}b_2+
\label{ea4}\ee
$$
+\left[\phi'^2-{C_1'\over C_1}{C_2'\over C_2}-{C_1'\over C_1}{C_3'\over C_3}-{C_2'\over C_2}{C_3'\over C_3}+\left(2{C_1'\over C_1}-{C_2'\over C_2}-{C_3'\over C_3}\right){D'\over D}+\left(2-{C_2\over C_3}-{C_3\over C_2}-2{Z_2A_2^2\over C_3}\right){k^2B\over C_1}\right]{g_1\over 3}=0
$$

Finally from the perturbations of the ($x_2x_2$)~ ($x_2x_3$)~ ($x_3x_3$) Einstein equations we obtain
\be
g_2''-\left({C_2'\over C_2}+{C_3'\over C_3}+{B'\over B}-{C_1'\over C_1}-{D'\over D}\right){g_2'\over 2}+{2ik\omega B(C_2-C_3)g_1\over C_1D}
+2Z_2\left(A_2'b_2'-{k^2A_2B b_2\over C_1}\right)+
\label{ea5}\ee
$$
+\left[\phi'^2-{C_1'\over C_1}{C_2'\over C_2}-{C_1'\over C_1}{C_3'\over C_3}+2{C_2'\over C_2}{C_3'\over C_3}-\left({C_1'\over C_1}+{C_2'\over C_2}+{C_3'\over C_3}\right){D'\over D}+3\omega^2{B\over D}-\right.
$$
$$
\left.-\left(1+{C_2\over C_3}+{C_3\over C_2}+{Z_2A_2^2\over 2C_3}\right){4k^2B\over C_1}\right]{g_2\over 3}=0
$$

By differentiating (\ref{ea3}) and using the other equations to remove the second derivatives of the fluctuations we obtain a new equation with first order derivatives that is equivalent to (\ref{ea3}).


\end{document}